\documentclass{article}

\usepackage{PRIMEarxiv}

\usepackage[utf8]{inputenc} 
\usepackage[T1]{fontenc}    
\usepackage{hyperref}       
\usepackage{url}            
\usepackage{csquotes}
\usepackage{booktabs}       
\usepackage{amsfonts}       
\usepackage{nicefrac}       
\usepackage{microtype}      
\usepackage{lipsum}
\usepackage{fancyhdr}       
\usepackage{graphicx}       
\graphicspath{{media/}}     

\title{Do We Run How We Say We Run? Formalization and Practice of Governance in OSS Communities
}

\author{
  Mahasweta Chakraborti \\
  University of California, Davis \\
  \texttt{mchakraborti@ucdavis.edu} \\
  \And
  Curtis Atkisson \\
  University of Massachusetts Amherst\\
  \texttt{catkisson@umass.edu}
  \And
  Ștefan Stănciulescu \\
  University of California, Davis \\
  \texttt{sstanciulescu@ucdavis.edu}
  \And
  Vladimir Filkov \\
  University of California, Davis \\
  \texttt{vfilkov@ucdavis.edu}
  \And
  Seth Frey \\
  University of California, Davis \\
  \texttt{sethfrey@ucdavis.edu}} 

\begin{document}
\maketitle

\begin{abstract}
Open Source Software (OSS) communities often resist regulation typical of traditional organizations. Yet formal governance systems are being increasingly adopted among communities, particularly through non-profit mentor foundations. Our study looks at the Apache Software Foundation Incubator program and 208 projects it supports. We assemble a scalable, semantic pipeline to discover and analyze the governance behavior of projects from their mailing lists. We then investigate the reception of formal policies among communities, through their own governance priorities and internalization of the policies. Our findings indicate that while communities observe formal requirements and policies as extensively as they are defined, their day-to-day governance focus does not dwell on topics that see most formal policy-making. Moreover formalization, be it dedicating governance focus or adopting policy, has limited association with project sustenance. 
\end{abstract}


\section{Introduction}
An exemplary instance of online peer production \cite{benkler2006wealth}, Open Source Software (OSS) has emerged as a multi-billion dollar informal industry supporting major contemporary tech enterprises, academia, and scientific research and development.  Over the past three decades, the increasing stakes of OSS have paved the way for several non-profit OSS foundations providing standardized project support and governance frameworks to hundreds of projects, notable among them, the Apache Software Foundation (ASF). These organizations serve OSS projects by providing mentoring, much-needed infrastructure (servers, centralized storage \cite{yin2021apache}, etc.),  legal aid around OSS licensing\cite{o2003guarding}, and well-maintained technical support \cite{Feller_Fitzgerald_Hissam_2007}. OSS foundations like the ASF have brought OSS into the mainstream, attracting large numbers of contributors and financial support \cite{Midha_Bhattacherjee_2012}.  

OSS  projects have often benefited from some degree of overarching coordination and governance \cite{Lee_Cole_2003, Sawhney_Prandelli_2000}.  Several of these foundations implement their own governance to manage projects and the developers they mentor.  Written, well-laid-out \emph{formal} policies steer and synchronize community operations, thus minimizing the costs of coordination and management~\cite{Butler_Joyce_Pike_2008,frey2022governing}.  At the same time, communities have often observed their own \emph{informal} rules and normative codes to structure activities, assign responsibilities, utilize project resources, and ensure sustained development \cite{Markus_2007,yan2023github, crowston2005coordination, Midha_Bhattacherjee_2012, Mockus_Fielding_Herbsleb_2002, Hippel_Krogh_2003, Lee_Cole_2003, heckman2007emergent}.  Consequently, community governance within foundation-mentored projects is a product of the foundation's policies, the project's own specific practices, and any interactions between those two sources of institutional structure. Hence, even among OSS projects from the same foundation, their decisions, actions, and ensuing interactions may reflect varied degrees of involvement with the centralized governance, as they may prefer to manage their community in their own fashion. 

Non-profit OSS foundations are steadily rising, with one survey finding 101 active organizations that host over 1,600 OSS projects as of 2018~\cite{izquierdo2018role, Izquierdo_Cabot_2020}. With mentored projects generally showing higher survival rates over independent communities \cite{Schweik_English_2012,yin2022open}, they are being increasingly viewed as a model to raise thriving projects producing usable, compliant software.  Yet, OSS governance is not without its quirks and challenges \cite{Lee_Cole_2003, crowston2005coordination, Shah_2006}. While foundations may bolster communities with resources and support,  the implications of such formalization for OSS has recently drawn significant research interest. Indeed, there have been instances where formal governance has produced little impact or has actually limited community flexibility and autonomy \cite{o2003guarding,Shah_2006,jensen_governance_2010}.  Hence, to assess the contribution of foundations towards OSS sustainability, we need to examine how they structure the mentored communities. We particularly look at how foundation policies are received in communities, reflected through operations as well as how they determine a project's governance focus.  

The Apache Software Foundation Incubator (ASFI) was founded by the Apache Software Foundation (ASF) in 2002, in part to propagate Apache's approach to OSS governance, and has mentored over 300 projects ('podlings') since. Several non-profits require interested projects to undergo initiation through an incubation program to learn the ways and requirements of the foundation. ASFI also evaluates projects for performance and overall organizational fit throughout their incubation, before accepting ('graduating') them for continued support, or 'retiring' them from the foundation. Being cognizant of the importance of project self-governance, ASFI empowers every project \cite{apacheself} to oversee its own governance with Podling Project Management Committees (PPMC). These PPMCs act as the interface between project developers and the ASF. The foundation's commitment to project self-reliance raises a fundamental question: what is the relationship of each project's emergent governance structure to the formal policies representing governance across the foundation?

Our study focuses on community-level governance among mentored projects and how they relate to foundation-level policies. We leverage developer conversations from ASFI's public mailing lists. Compared to traditional approaches like surveys, interviews, or other forms of qualitative inference, retrieving behavioral measures from trace data is faster, convenient for replication across foundations, and less susceptible to reporting bias while offering more granular, real-time insight. We assess each project's governance efforts and resulting operational structuring through the routinized \emph{governed activities} they perform.  Next, we evaluate their policy \emph{internalization}, i.e. the extent to which ASFI formal policies structure their community governance and frame their governed activities. We analyze how the extent of community governance efforts and policy internalization relate to ASFI's extent of regulation (number of rules) across different governance topics.  Finally, we empirically investigate how community governance and its extent of formal policy internalization together explain its Incubator outcomes. Our contributions and findings are as follows:

\begin{enumerate}
\item We demonstrate a scalable approach, based on semi-supervised learning, to understand governance across peer-production communities, both its formal specification and lived instantiation. 
\item A foundation-level analysis of ASFI projects shows that the extent of policy regulation --- the number of rules structuring different governance topics --- is not mirrored in practice through the extent of governed activity. Yet  governed activity tends to be framed by policies in topics where they are extensively defined, as indicated through policy internalization among projects. Therefore, while communities show greater acknowledgment of formal policies in the topics where they are extensively laid out by the ASFI, such topics do not necessarily elicit more governance efforts from communities.
\item When it comes to sustaining the community and efficient development towards graduation, dedicating governance focus or internalizing policies from topics highly regulated/prioritized in formal policies had little association with the odds of success. All in all, formalized policies in OSS communities may not accurately reflect their underlying patterns of governance.
\end{enumerate}

\section{Related Work} \label{review}
Open Source governance includes all organizational structures and coordination mechanisms that regulate community interactions as well as product development. Prior work has extensively explored OSS community governance in terms of decision-making \cite{heckman2007emergent, yin2022open}, assignment of tasks \cite{crowston2005coordination, Midha_Bhattacherjee_2012}, managing developer roles and access \cite{Mockus_Fielding_Herbsleb_2002, Hippel_Krogh_2003}, mentorship~\cite{atkisson2022mentors}, code quality, review, and contribution \cite{Lee_Cole_2003,10.1145/3540250.3549132}, etc. 

Community governance has been treated as an expansive, multi-level system of mutually interactive socio-technical networks \cite{jensen_governance_2010,frey2019emergence}. Meanwhile, Schweik et al. studied OSS projects at scale on SourceForge and found governance structures to be generally informal and lean, with increased sophistication and formal rules as communities grew \cite{schweik2007tragedy}, Similar findings are also echoed by O'Mahoney's work on the Debian Linux community's evolving governance~\cite{o2007governance}. Community-level analysis of Apache Incubator projects also found that more successful projects showed greater adoption and use of definitive rules and norms \cite{yin2022open}. Heckmann et al's investigation of decision-making processes further found that in well-performing projects developers and users participated more proactively in steering the course of the project \cite{heckman2007emergent}. 

Leadership is a crucial aspect of OSS governance, where developers with greater technical initiative, development prowess, and effective communication strategies generally emerge to fill administrative roles \cite{hergueux_follow_2022}. Analysis of decision episodes in communities found administrators to be critical drivers during the initial phases of a project \cite{heckman2007emergent}. Meanwhile, Atkisson specifically examined individual mentors of the Apache Incubator and found a significant correlation between who managed a project and its odds of graduation \cite{atkisson_mentors_2022}. Investigation of communities on SourceForge found that while a sizeable fraction (around 15-20\%) of successful projects comprised a stable community with dedicated users, the rest showed rapid growth and were often led by a 'benevolent dictator' \cite{Schweik_English_2013,Schweik_English_2012}.


Prior work has explored the challenges of OSS moderation. Attempts towards greater inclusiveness by enforcing community codes of conduct (CoC's) have often received limited engagement or been perceived as distractions from core development priorities \cite{li_code_2021}.  Several studies have focused on interactions within foundation-led communities.  A qualitative cost-benefit analysis of Apache Incubator policies found that the implementation efforts and payoffs are evenly balanced between projects and the ASF \cite{Sen_Atkisson_Schweik_2022}.  The implications of congruence/dissonance become particularly salient when it concerns software licensing. The rigor of the licensing requirements, including ASF's rights over individual contributions, has often seen varied reception and interpretation among OSS developers \cite{o2003guarding}. Sun's introduction of changes in the Netbeans licensing scheme threatened the collapse of the very project \cite{jensen_governance_2010}. Stringent terms set by corporations supporting gated OSS communities often turned away sincere contributors or restricted usage of the product, thus hindering developer engagement and community health \cite{Shah_2006}. 

While prior work has either focused on foundations or community dynamics, a limited number have empirically treated their mutual interactions unraveling  in real-time \cite{ostrom2007going,yin2022open}. Moreover, they have generally focused on a particular aspect of governance, such as licensing, through case studies of a select number of projects. We attempt to capture the multifacetedness of OSS governance (including but not limited to licensing, trademarks, documentation, committees, voting, etc.) and study hundreds of mentored projects. Motivated by collective action theory and behavior in communities of practice, we proceed to investigate the governance behavior of OSS communities around formalization.


\section{Theoretical Motivation}

\subsection{Institutional Theory}
\label{section:institheory}
OSS communities, generally comprising transient volunteer developers centered around a core of long-term contributors, organize in a decentralized fashion to create software for open use and distribution. This phenomenon has been framed in terms of the peer production of public goods, making OSS communities an increasingly important locus of online collective action research \cite{benkler2006wealth}. 

Institutions are defined as “… prescriptions that humans use to organize all forms of repetitive and structured interactions …” \cite{ostrom2009understanding}. For a collectively maintained resource such as an OSS community, governance includes all formal and informal rules for management and production, along with the mechanisms for such policy design, reform, and implementation. \cite{Schweik_English_2012, McGinnis_2011}. 


OSS governance lies on the spectrum between purely self-interest-driven spontaneous governance ("the invisible hand") and intentional governance~\cite{de2007governance}. Polycentric governance refers to a condition where there are overlapping interests between multiple centers of authority \cite{McGinnis_2011,mcginnis1999polycentricity,jhaver2021decentralizing}. This often implies varying degrees of interdependence and autonomy among concurrent governments. For example, while ASFI encourages projects to admit consistent contributors, the specific process of admission is left to each project  community itself~\cite{Sen_Atkisson_Schweik_2022,yin2021apache}. The dynamic nature of organizational fit is especially evident in decentralized\cite{de2007governance,weick1976educational}, ideology-rich environments like OSS projects~\cite{stewart2006impact}, notably as resource abundance varies~\cite{scott2005institutional}. It is our goal in this paper to study the extent of internalization of ASFI's governance in regular project operations, along with the different themes of its rules and policies, across graduated and retired projects.


\subsection{Communities of Practice and Organizational Learning}
\label{section:orgtheory}
OSS projects are essentially online communities of practice \cite{Lave_Wenger_1991, Brown_Duguid_1991, orr1990talking}, where coordinated operations are studied in terms of routines. Routines stem from beliefs, cognitive scripts, habitual conventions as well as evolving norms as they translate into 'repeated patterns of actions' across appropriate settings \cite{Cohen_Burkhart_Dosi_Egidi_Marengo_Warglien_Winter_1996, Levitt_March_1988}. These include management, standard operating procedures e.g. workflows, or experiential strategies encoded into everyday activities and associated interactions \cite{Levitt_March_1988,cyert1963behavioral,winter1982evolutionary}. Community routines may not be only technical, and may also emerge to coordinate developers through informal norms and social control \cite{Markus_2007,Mockus_Fielding_Herbsleb_2002, Hippel_Krogh_2003}.  For example, developers use their particular routines for managing and deploying builds, incorporating patches, testing, prioritizing issues, et cetera. Similarly, communities also perform a sequence of routines when it comes to more formal events like setting up committees, organizing conferences, and ratifying releases. 

Routines are generally stable \cite{Cohen_Bacdayan_1994}, until changes in organization, technology, development goals, or other events cause them to evolve \cite{Feldman_2003, Pentland_2005, Feldman_2000}. OSS projects are dynamic and decentralized with fluid membership \cite{Vasilescu_Filkov_Serebrenik_2015} and may thus be inventive and flexible in their norms \cite{meyer1977institutionalized,anthony1984constitution}. Consider the following email from Apache Netbeans dated 9/13/2017. 
ASFI does not cover code management. Yet their projects themselves usually chose between two approaches: review-then-commit (RTC) and commit-then-review (CTR). The example shows deliberation among Netbeans developers on their appropriateness and scope: 

\begin{displayquote}
different asf projects have different policies. the important part is that we should have a common understanding about our commit policy. there might e.g. be a branch for the next release where rtc (review then commit) is applied. that's useful when preparing a release or for maintenance releases we still actively maintain. and beside that we might have a 'future' branch (e.g. on master) or multiple feature branches where ctr (commit then review) is standard. most asf projects have the whole repo on ctr...
\end{displayquote}

Incubator policies are set up through pragmatic planning. The  observed influence of the foundation's policies on a mentored project's routine operations indicates how its governance has been internalized in the community. The more community members discuss and describe activity in a way that resembles the framed policy, the more we can argue that members have internalized the formal description. At the same time, through learning and discovery \cite{March_1991}, communities may also prefer procedures and protocols when Incubator policies are deemed less effective, inadequately defined \cite{meyer1977institutionalized,weick1976educational} or fall short of their needs. This further motivates us to understand the impact of foundations on projects through policy internalization across sustained community practices.

\label{section:RRQ}
\section{Research Questions}

Formal rules and policies are critical in shaping the basic structure and guiding activity in an organization \cite{Pentland_2005,anthony1984constitution}. Foundation Incubators implement systematic policies to coordinate and promote community engagement and productivity. These establish baseline standards and rules for participation across all the diverse member projects, may define certain roles and offices for leadership, assign responsibilities, as well as lay out the scope of various activities.   At the same time, routines also reflect the project community's own implicit governance, i.e. informal beliefs, norms, codes of conduct, and other practices. Therefore, polycentric governance in foundation projects stems from governance among individual communities (Project Management committees (PMCs), as well as all other informal rules and developer norms) alongside the ASFI itself. Situated in the backdrop of OSS-foundation polycentricity, this section presents our research questions which look at community governance and policy internalization across the different aspects of ASFI governance.


The formalization of governance in traditionally volunteer-driven communities has been a contentious theme. OSS pioneer Eric Raymond observed that the “number of hoops” or too many formalized procedures and rules may drive away potential skilled contributors \cite{Raymond_1999,schweik2007tragedy}. Extensive regulation may introduce additional requirements and necessitate the enactment of institutional obligations. Therefore, communities may be expected to show more governed activity in domains that are heavily policied, given their presumed importance in the ASFI ecosystem. As a result, we may expect a positive relation between the number of policies and the frequency of observed routine activities in a particular area of governance. 

While there are concerns about redundant routines and overheads, lack of regulation may cause individuals/communities to draw upon larger social and cultural constructs for predictability. Such "tyranny of structurelessness" may perpetuate broader social inequalities~\cite{Freeman_1972}. The idea of "green tape" encapsulates the potential of policy to provide clarity and certainty, focus organizational attention, and convey legitimacy~\cite{DeHart-Davis_Chen_Little_2013}. Implications may also extend to OSS formalization, whereby extensive yet well-designed policies may streamline rather than divert developer efforts. However, in domains where regulatory clarity is limited, greater project activity may become necessary to sustain development. 

RQ1 explores how the extent of policy-making relates to the governance priorities and operations among mentored projects. We identify governance concerns/topics actively shared between the ASFI and its projects, through policy documents and extensive mailing lists across 208 communities. Since structuration from the mutual interaction of foundation policies and community governance determines the routine behavior of projects, we aggregate all similar activities from email conversations and examine their correlation with the topical distribution of ASFI policies. 

\textbf{RQ1:} How does Incubator regulation relate to community-level governed activities across different governance topics? 

Institutions manifest through the practice of routines formalized by such established rules\cite{Lammers_2011}.  As mentored projects increasingly internalize foundation policies, their operations are expected to be generally constrained and enacted through routines prescribed by such rules. Yet, community governance also requires the dynamic selection and adaptation of various other routines (Section \ref{section:orgtheory} ). Therefore, we may expect variation in the influence of ASFI policy on governed activity, along the different governance concerns. 

Well-designed rules seek to reduce uncertainty and can act as formulaic precedents to replicate success across mentored projects \cite{meyer1977institutionalized}, or at least help standardize the provision for Incubator resources. Therefore, extensive regulation in a certain area of policy-making (i.e., more rules outlining a wide range of organizational possibilities), may induce greater adoption if it facilitates project functioning and improves efficiency. 

On the other hand, activities and related exchanges in a topic may deliberate policy to only an extent, while their actual operations may reflect a marked departure from formal structure \cite{weick1976educational,DiMaggio_Powell_1983}. This may be especially true when certain institutional obligations are ceremonial or necessary to maintain affiliation with the ASFI but are less relevant in day-to-day development. If such is the case, the observable policy internalization among communities across different governance topics may not be correlated to the extent of policy overseeing the topic.


We might expect alignment between the amount of formal policy on a topic and how resulting policy prescriptions are internalized in practice. Organizations engage in many functions, some of which are more critical than others. More important functions may be marked by a greater amount of policy formalizing behavior and may elicit greater internalization, toward more compliant execution. On the other hand, if policy extent is driven more by the complexity than the criticality of a governance subject, then that complexity may paradoxically predict a greater quantity of policy, for its various cases, and also less internalization, as practitioners take license from that very complexity to exercise greater discretion in how they execute. 

RQ2 explores how the extent of policy-making relates to the formal policy internalization among projects.  For all topical governed activities we measure policy internalization in terms of how discourse about those activities in general semantically reflects the policies formalizing those activities. Finally, we examine how such internalization varies with the extent of regulation across topics.

\textbf{RQ2:} How do the levels of policy internalization in governed activities relate to ASFI policy extent across different topics?

For an Incubator program to realize its goals, it is important to assess the association between its governance and project outcomes. At the same time, it becomes equally important for aspiring communities to understand behavior associated with communities that succeed in Incubator programs, particularly the extent of community governance as well the impact of foundation governance on such operations. 

ASFI lays down three primary criteria to determine if a project has potential and is capable of sustaining development: 1) there is community activity evidenced by at least two releases, 2) the releases are compliant with the Apache license, and 3) the committers of a project are drawn from at least three entities (companies, research groups, etc.)~\cite{ASF2023}. The remainder of the policies serve to help the project achieve those goals.

While RQ1 and RQ2 measure \emph{if} there is a relationship between formal policy and community governance, RQ3 uses an externally valid measure of project outcomes to determine whether there \emph{should be} a relationship i.e. whether communities align governance focus or internalize policies in topics with more formal rules, in order to successfully realize their objectives. In particular, it examines if community governance efforts or the adoption of policies around formalization correlates to their graduation odds in the ASFI. 

We pursue RQ3 through a project-level regression of all governed activities (frequency of structured, routine operations) among individual projects alongside the policy internalization among such operations (semantic similarity of governed activities to policies) against a binary measure of project success (graduation/retirement from the Incubator).

\textbf{RQ3:} How do governed activities and the extent of policy internalization relate to the success of projects?


\section{Data and Methods} 

\subsection{Variables of Interest}
\subsubsection{Governance Measures:} We pursue two discursive measures of community governance from developer conversations in mailing lists, namely all governed activity and their internalization of Incubator policies.  Traditionally public and open access, OSS mailing lists are key to collaboration as they promote transparent peer review \cite{Lee_Cole_2003} and solicit reciprocal contributions \cite{o2003guarding}. Unlike issue tracking and version control logs, these also contain exchanges beyond technical development, such as product planning, community management, ratification of major decisions, licensing, etc. Further, due to explicit ASF policies, all project activity are comprehensively archived across public mailing lists (“If it didn’t happen on the mailing list, it didn’t happen” \cite{yin2021apache}).  

Prior work has extensively used organizational communications for understanding participant behavior and performance, including 0SS \cite{jensen_governance_2010,yin2022open,hergueux_follow_2022}. Li et al. used a grounded theoretic approach to understand the adoption and reception of community codes of conduct from developer exchanges. Affective features in developer messages have been used to predict leadership qualities among OSS developers \cite{hergueux_follow_2022}, while Srivastava et al. studied enculturation and employee exit, where they treated individual's linguistic divergence as a measure of cultural fit \cite{srivastava_enculturation_2018}. 

We described in Sec. \ref{section:orgtheory} how routines reflect all prevailing governing norms among projects. We first identify the different governance concerns shared between projects and the Incubator by means of topic modeling of policies and conversations, and represent the following two measures by project and governance topic:

\emph{Governed Activity:} The total number of recurring or routine activities about a governance topic, as discussed in a project's mailing list. Higher presence of governed activity indicates greater governance efforts to structure and routinize community operations. For example, if a community establishes a norm for ratifying releases, future releases will likely follow the established schema. In ASFI projects, such governance is a culmination of the foundation's policies as well as the underlying codes and norms of the community developers. Recurring activities are aggregated over their textual similarity. 

\emph{Policy Internalization}: This measure represents the extent to which governed activities are structured by ASFI policies. Therefore, higher policy internalization in governed activities indicates greater integration of the foundation into the community's governance.

Methods explored to operationalize internalization included evaluating direct compliance/entailment between an observed activity and policies. Such binary measurements were found to be insufficient to account for the drift between formally articulated statements (framed policies) and informal, practical discourse (conversations) or importantly, reflect graded changes along the rates of institutional diffusion \cite{Strang_Meyer_1993}. For example, observations from initial developer discussions over a release vote, an ASF-specific requirement, to an actual voting event are important to understand the gradual internalization of governing institutions. 

For a topical governed activity in a project, we measure policy internalization through its semantic similarity against policies within the respective topic. Semantic similarity is an assessment of meaningful and conceptual relationships between texts \cite{jurafsky2000speech}. Measured on a continuous [0,1] scale, semantic similarity rates text pairs higher (lower) values for agreement (contradiction) \cite{cer2017semeval,wieting2020bilingual}. Moreover, semantic similarity can be used to quantify activities that are neutral but indicate institutional diffusion, through their degree of resemblance in how they invoke roles, designated responsibilities, and requirements outlined by a policy. 


\emph{Policy Extent:} A foundation-level variable indicating the extent of ASFI's regulation across topics. It is represented as the frequency (count) of formal rules overseeing each governance topic, with higher values (number of rules) in a topic indicating greater ASFI regulation. 

\subsubsection{Project membership and activity}: Projects in ASFI are diverse, and their governance and Incubator outcome may also be subject to community structure, activity levels, etc. Since we are interested in analyzing how governance behavior correlates to project sustainability, our analysis has to simultaneously control for project attributes, such as community size and development intensity. We incorporate four suitable covariates in our analysis through \textit{community size (committers), number of commits, code base size (lines of code; LOC)}, and finally the frequency of interaction among the project developers (\textit{developer emails}) over project mailing lists.

\subsection{Datasets}
We center our analysis of ASFI governance through a set of 234 comprehensive policies which were coded across the key ASFI documents and guidelines \cite{Sen_Atkisson_Schweik_2022}. These span multiple sources such as the official Apache Incubator policy manual, the community guide, the Podling Project Management Committee (PPMC) guide, the Apache cookbook, the mentorship guide, the graduation and retirement guides, and finally the release management guide. 

In the ASFI, project incubation lasts up to several months followed by an assessment and a formal vote to decide on graduation into ASF for continued support or retirement. Yin et al. scraped all mailing lists across 269 Apache projects from when they joined the Incubator and up to their last day in the ASFI \cite{yin2021apache}. Since we solely focus on norms and activities within communities, we only retain the ‘dev’ (community developers) subdirectory emails across all projects. We exclude redundant content such as auto-generated emails, for issues posted and resolved, and other development-related notifications (JIRA, Github) through source address-based filtering. Periodic emails were also circulated by the Incubator Project Management Committees (IPMC) or project mentors, which were formal, administrative, and generally concerned progress reporting. All such emails have a fixed format and were identified and filtered through string matching. This mitigates potential bias in measurements from to superfluous policy content from the administration, as our subsequent analysis concerns governance-related behavior within and among community developers only.

For project-level covariates, we obtain commits, lines of code, and the number of active contributors.  ASFI projects use GitHub, Subversion, or a combination of both for maintaining their codebase. Stănciulescu et al. \,\cite{stanciulescu2022code} extracted monthly performance metrics for 218 ASFI projects  their incubtion. However, the tooling infrastructure they developed only supported mining software metrics from Git repositories. Moreover, Yin et al. mined project mailing lists up to Jan 2021, including ones that were mostly SVN-based, while Stănciulescu et al. span projects from March 2003 up to May 2021. Given these differences, we based our study only on those projects that are common to both datasets. This yielded 214 projects for which both project measures and email data were available. 

Moreover, there were some differences in the way these data were collected. Yin collected data in time windows of 30 days, whereas the other dataset collected data on a monthly basis (calendar timestamps). To resolve this mismatch, we modified the collection timeline to a 30 days time window in the tool provided by Stănciulescu et al, to match the time window in the dataset from Yin et al., and repeated the measurements for our variables of interest for these 214 projects.

\subsection{Measurements}

\label{section:method}
\begin{figure}[!h]
\centering
\frame{
\includegraphics[width=.95\textwidth,height=0.3\paperheight]{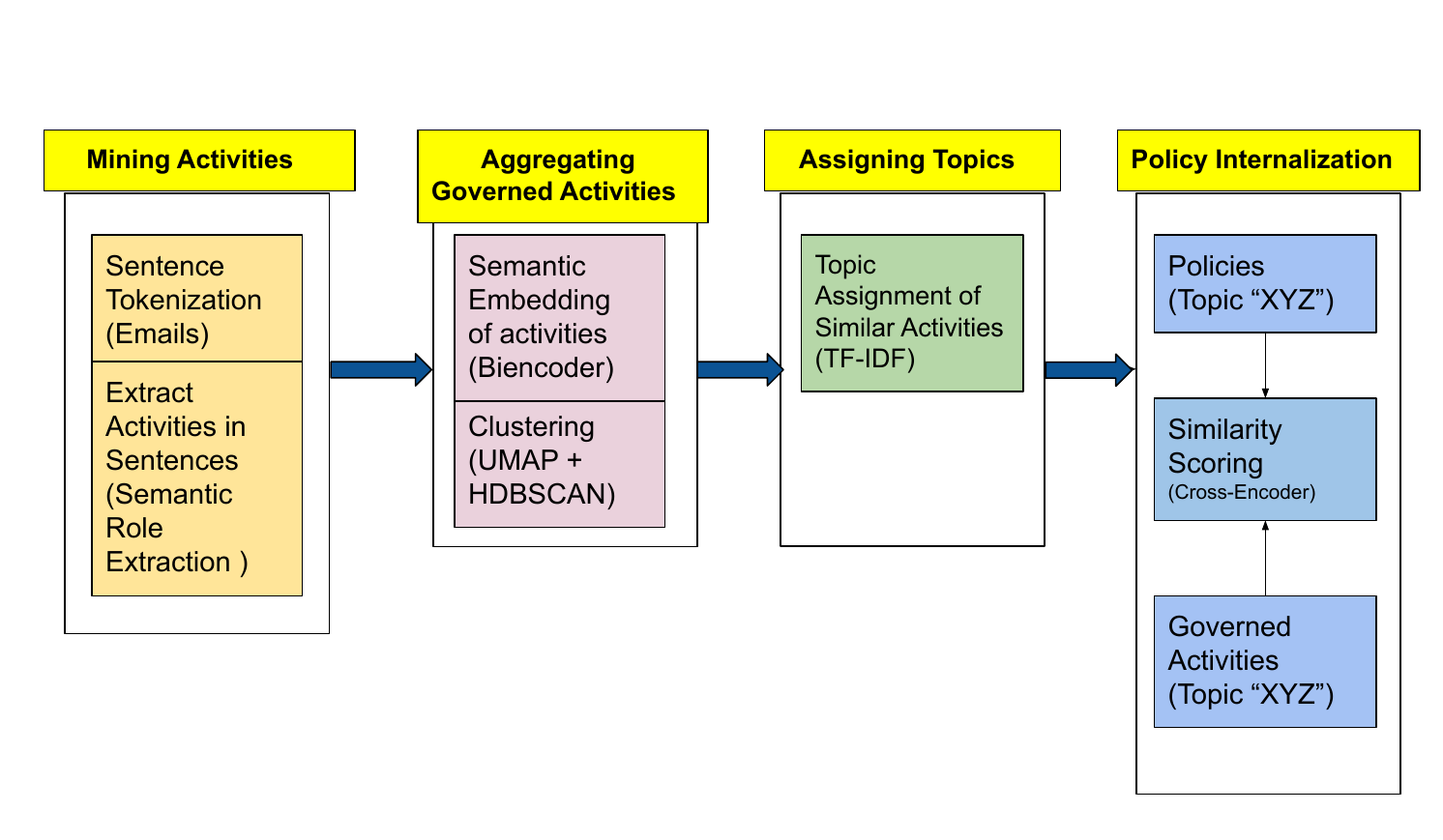}}
\caption{Language modeling pipeline for extracting activities, aggregating routine governed behavior, and evaluating internalization. }
\end{figure}

\subsubsection{Extracting activities}

Routines have been studied at multiple levels, from the most nuclear activities to complete processes. The most fundamental unit, the performance program \cite{Pentland_1995,march1958organizations} is defined as a 'chunk' of scripted activity, generally a routine in itself or part of a larger process.  To capture organizational routines from ASF email discourse, email texts and policies were first tokenized into sentences through StanfordNLP’s Stanza library \cite{qi2020stanza}. We next turn our attention to extracting different activities from within these sentences. 

This serves several purposes. Firstly, most existing language models, including ones subsequently used, encounter complexity overheads and truncate long sentence inputs beyond a token length. Secondly, sentences can be compound, conveying multiple activities with their specific context, and possibly spanning different topics (Table \ref{tab:srl_ex}). Therefore decomposing sentences into granular units of analysis like performance programs allows depth and insight in subsequent analysis. 
 
We decompose sentences while preserving their context. Context is important in understanding different routines and their place in the development ecosystem (E.g. \textit{`Projects \textbf{requesting} Apache infrastructure'} vs. \textit{`Project Management Committee \textbf{requesting} progress report'} or \textit{`Projects \textbf{issuing} press release'} vs. \textit{`Resolve \textbf{issues} that are release blockers'}). To attain fine-grained extraction of different activities and their context nested within sentences, we use semantic role labeling. 

Semantic role labeling or SRL \cite{jurafsky2000speech} is an NLP task that extracts roles (actors, direct or indirect objects, etc.) associated with an action (verb) along with other modifiers from a sentence. Additionally, SRL also extracts constituents with contextual information such as the time of act, manner, direction, goal, purpose, cause, etc \cite{bonial2010propbank}.

\begin{table}[!h]
\centering


\noindent\fbox{%
 \parbox{0.95\columnwidth}{%
\textbf{Original Policy:} \\ 'After a vote has finished, the ipmc must send a notice email to the board and then wait for 72 hours before inviting the proposed member' \\

\textbf{Semantic Role Parsing:} \\
'ARG0': ['the ipmc'], 'ARGM-MOD': ['must'], 'V': ['send'], 'ARG1': ['a notice'], 'ARGM-DIR': ['email'], 'ARG2': ['to the board'], 'ARGM-TMP': ['after a vote has finished']

'ARG1': ['the ipmc'], 'ARGM-MOD': ['must'], 'V': ['wait'], 'ARGM-TMP': ['after a vote has finished', 'then', 'for 72 hours', 'before inviting the proposed member'] \\

\textbf{Performance Programs (After reconstitution):} \\
'After a vote has finished the ipmc must send a notice email to the board'\\
'After a vote has finished the ipmc must then wait for 72 hours before inviting the proposed member'\\
 }}%
\caption{\small Activities from compound sentences through Semantic Role Labeling (SRL). ARG0 denotes agent, ARG1-ARG5 are direct/indirect objects, ARG-MOD indicate modals while ARG-TMP and ARG-DIR are the temporal and directional arguments respectively} 
\label{tab:srl_email}
\end{table}

\begin{table}[!h]
\centering


\noindent\fbox{%
 \parbox{0.95\columnwidth}{%

\textbf{Original Sentence:} \\ '( 1 ) I'll be away from my computer starting Friday and through the New Year, so I won't be able to do much to help if folks want to release 2.1 during that time ( not even testing ).'  (Apache Roller, 12/21/2005)\\

\textbf{After SRL and reconstitution:} \\
'I'll be away from my computer starting Friday and through the New Year' (Schedules/Events) \\

I won't be able to do much to help if folks want to release 2.1 during that time ( not even testing )' \\ (Release Management)\\ }}%
\caption{\small Capturing granularity: Sentences spanning multiple, thematically distinct operations. In this example, a developer shares their vacation timeline to the community in general, while also discussing implications for a tentative release. Topics indicated for each activity are inferred as described in Section \ref{topic}} 
\label{tab:srl_ex}
\end{table}

We chose a BERT \cite{devlin2018bert} based implementation of SRL \cite{shi2019simple} developed by AllenNLP on the Propbank annotation scheme. The model holds a state-of-the-art performance on the English Propbank (Newswire) as well as a test F1 score of 0.864 on the Ontonotes 5.0 dataset. We identify all possible semantic roles associated with each distinct verb from compound sentences. These SRL frames were reconstituted into distinct activities, by reordering the semantic roles and all other contextual arguments for each verb, along with their relative positions from the original sentence. The 723,863 developer emails in our data generated 2,248,950 expressions of activities. 

In governance research, rules are specified in terms of grammatical constituents representing the governing (committees, boards, etc.), the governed (e.g. committers), the activities they undertake, and the conditions they entail (e.g. voting before a release) \cite{crawford1995grammar}.  Our policy reference data \cite{Sen_Atkisson_Schweik_2022} comprised descriptive policies spanning multiple nested rules (Table. \ref{tab:srl_email}).  Therefore, SRL-based preprocessing was also extended to the policy documents, whereby the 234 policy descriptions from Sen et al. were parsed into 422 individual rules. 

Finally, we conduct an additional pre-processing step. Developers often use mailing lists for technical discussions and clarifications. As a result, they often contain stack traces, logs, etc. which may be parsed as regular activities. We restrict our analysis to human-readable, standard English-language data, which can be compared and interpreted against governance policies such as those of ASFI. We detect and retain only English texts using a HuggingFace XLM-Roberta-base model \cite{conneau2019unsupervised} trained for language identification. This reduced the number of extracted activities to 2,029,691.

\subsubsection{Governed Activities: Aggregating routines}
As described in (Section~\ref{section:orgtheory}), routines are activities carried out time and again, under specific circumstances \cite{Cohen_Burkhart_Dosi_Egidi_Marengo_Warglien_Winter_1996}. Unlike well-documented formal policies, routines are more dynamic and span activities dictated by emerging norms and operational priorities. Hence, it is extremely challenging to comprehensively codify activities in a community and train models that can discriminate routine behavior from non-routine ones. 

Importantly, we are interested in a pipeline that supports governance analysis across diverse online communities. Since routines are influenced by technological trends, the nature of the product, the specific community, utilities involved, etc., there may arise inaccuracies from data when extending a supervised model specifically built on ASFI data, to other communities and foundations. Based on theoretical definitions of our construct of interest (i.e. governed activities are routine or 'recurring' operations), we leverage alternative learning methods compatible with our goals. We hereby describe our approach to discovering routines as similar activities in email data, through semi-supervised clustering. 

We find similar ('recurring') activities through semantic similarity-based aggregation \cite{Reimers_Gurevych_2019}. Popular approaches to semantic representations include word level \cite{mikolov2013efficient, pennington2014glove}, sentence level \cite{Conneau_Kiela_Schwenk_Barrault_Bordes_2017,cer2018universal, wieting2015towards}, and more recently language model-based approaches which allow for more advanced representation learning for different semantic tasks.  

The biencoder architecture was developed for computationally efficient semantic encoding of texts \cite{Reimers_Gurevych_2019}. They involve training a Siamese network of two identical, transformers to generate contextual encodings for two distinct text inputs. The averaged output from each transformer is then subjected to a cosine similarity loss objective function. By the end of the joint fine-tuning, both the transformers are capable of independently generating semantic embeddings for any given text input. Huggingface \cite{wolf2019huggingface} hosts multiple domain-specific biencoders. We use a general-purpose bi-encoder pre-trained on the domain-relevant corpus from Stack Overflow, a question-answer platform specially used by developers. All transformer-based experiments henceforth were conducted through a single Tesla T4 GPU.

Next, for aggregating encoded texts, we use BERTopic \cite{grootendorst2022bertopic}. It supports hierarchical density-based clustering or HDBSCAN \cite{mcinnes2017hdbscan} for most Hugginface binecoders, followed by topic modeling of the inferred clusters. To train the clustering model, we uniformly sample 100,000 activities out of all the 2,029,691 activities previously extracted. Modeling activities across projects together allows for identifying and grouping them under a set of shared governance topics.

To cluster community activities intersecting with ASFI concerns, the 422 rules from ASF policies are passed as initial seeds to BERTopic. For best clustering results, we conducted hyperparameter tuning for BERTopic's HDBSCAN through Density-based clustering validity or DBCV measures \cite{moulavi2014density}. DBCV scores rate density-based models from -1 to +1, with higher values indicating better clustering quality. To find hyperparameters returning maximum DBCV, we tune over the following  HDBSCAN arguments: minimum cluster size and minimum samples. Higher values of cluster size threshold might lead to the merging of clusters, while sample size promotes dense clustering and more outliers. Both parameters were varied in combinations from 10 (0.0001\% of sample size) to 100 (0.001\% of sample size). Prior to clustering, BERTopic also uses Uniform Manifold Approximation and Projection or UMAP for dimension reduction of embeddings. The number of neighbors parameter in UMAP decides the trade-off between preserving the global and local structure and was also varied between 10 and 100. We retain the model with the best relative DBCV score.  

\subsubsection{Topic modeling of governed activities} \label{topic}
BERTopic finally conducts TF-IDF across the dense clusters of governed activities to assign them topics. Words from the rules were used to suitably reweigh inverse document frequency of words. Topic coherence metrics \cite{Röder_Both_Hinneburg_2015} supported by Gensim \cite{vrehuuvrek2011gensim} evaluate topic modeling  performance on a scale of 0 to 1. Our final model shows a topic coherence $C_{v}$ of 0.683, indicating strong topic correlation. 

Policy documents often contain canonical descriptions of norms and processes, that are often dated and removed from practical operations \cite{Brown_Duguid_1991,orr1990talking}. After clustering, 106 out of the 422 policy rules were disregarded as outliers, due to negligible mention over emails. A total of 42 distinct topics were identified between ASFI policies and email activities, and 211 topic clusters were discovered among all activities. Around 493,008 activities were found to belong under these 42 governance topics from ASF. Final topic label assignments were deduced based on the assigned policies and top keywords from each topic, and overall domain knowledge of ASFI.

\subsubsection{Measuring institutional internalization}
For governed activities under any ASFI governance topic, we measure the extent to which they reflect ASFI policies overseeing the same topic. Cross-encoders or poly-encoders \cite{Reimers_Gurevych_2019} are a standard language model for semantic comparison. They treat sentences or text to be compared as simultaneous inputs and attend them jointly for semantic scoring. Biencoders and cross-encoders are often used together for information retrieval and text ranking. While biencoders can encode individual sentences to support high-level clustering over large sets of text, cross encoders are suitable for more precise, pairwise comparison between smaller sets of texts \cite{thakur2020augmented}. 

We use a Distil-RoBERTa base cross-encoder from Huggingface which rates text pairs on a continuous scale of 0 to 1, with higher scores indicating greater similarity The model demonstrated a Spearman rank correlation of 0.87 with respect to the human-annotated scores from the STS text similarity benchmark \cite{cer2017semeval}. Using this cross-encoder, we compare every governed activity against all the rules assigned to the same governance topic to find the ones it resembles most closely. The mutual semantic similarity score of the governed activity and the closest policy is used to represent the activity's extent of ASFI policy \emph{internalization}. Consequently, we obtain internalization scores for all the 493,008 governed activities under each of the 42 governance topics.


\subsection{Analysis}
RQ1 and RQ2 pursue an ASFI-level exploratory analysis of our governance measures along the policy extent. RQ1 compares the proportions of ASFI rules (level of regulation) and project-level governed activity across the topics, while RQ2 follows up by assessing the distribution of ASF policy internalization in activities. 


Finally, for RQ3, we examine governance behavior among projects, against their graduation or retirement from incubation. We fit a generalized logistic regression (GLM) binomial model of project-level measurements of governance as well as the covariates, against their respective incubation outcome. We conduct our analysis through the GLM suite (regression, multicollinearity check, and validation of assumptions) supported by the \textit{statsmodel} package in Python. LASSO-based variable selection is conducted prior to regression and inference, for which we use the \textit{group-lasso} Python package. We set the significance level of our analysis at the standard $ p < 0.05 $.

\begin{figure*}[t!] 
\frame{\includegraphics[width=.9\textwidth,height=0.5\paperheight]{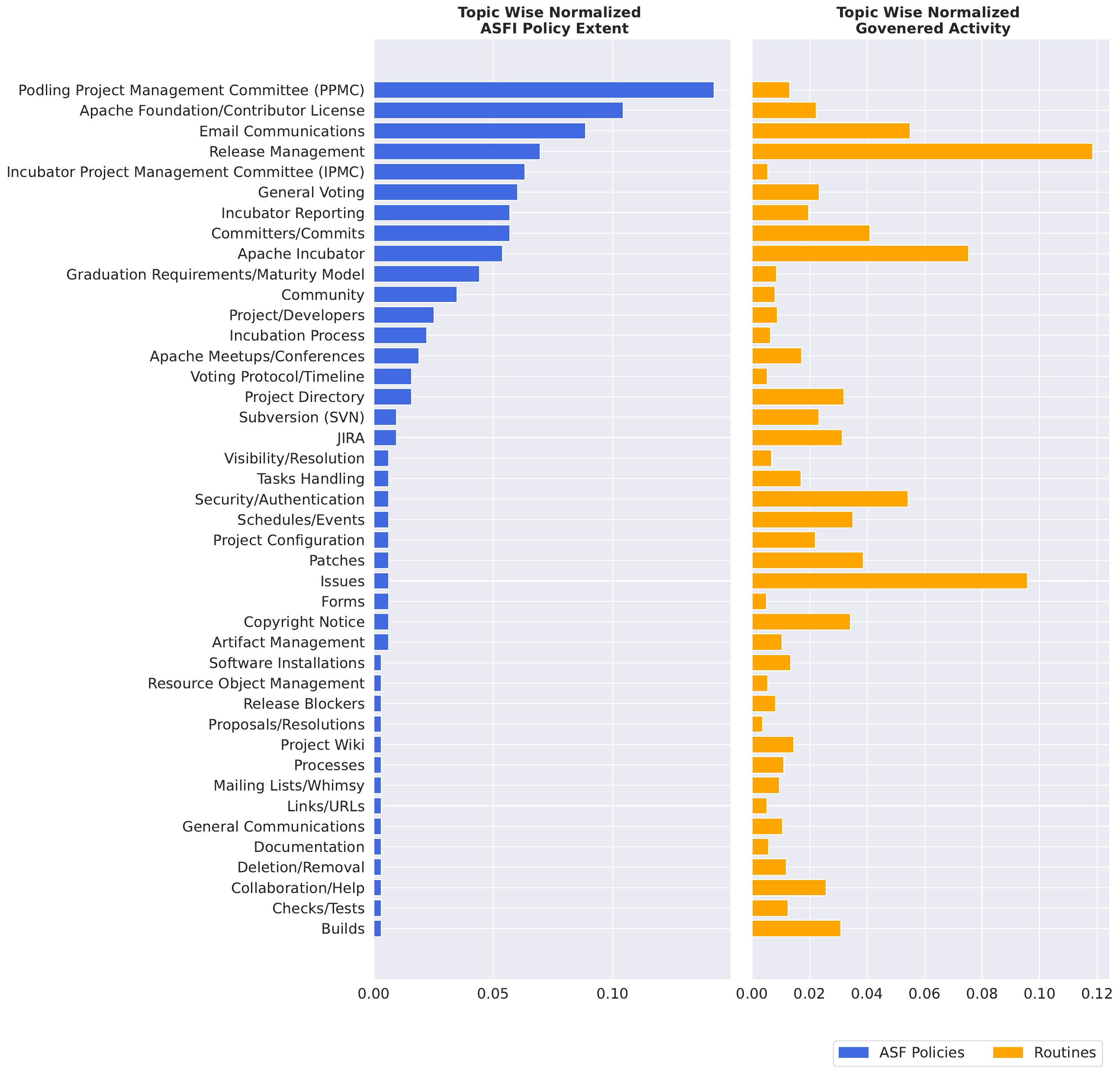}}
\caption{Left: Distribution of ASFI policy extent across governance topics. Right: Distribution of governed activity of projects across different governance topics. Governed activity was not found to be significantly correlated to policy extent.}
\label{fig:RQ1}
\end{figure*}
\subsection{RQ1: How does Incubator regulation relate to community-level governed activities across different governance topics?}
As described in (Section~\ref{section:institheory}), we focus our analysis on governance topics shared between the ASFI and its mentored projects. We visualize ASF's policy extent against the distribution of governed activity along topics (Figure~\ref{fig:RQ1}). A Pearson correlation test between the distributions was found to be 0.23 (\emph{p = 0.13}), indicating that how communities perform governed activities across topics is uncorrelated with the amount of policy structuring those topics.

\begin{figure}[t!] 
\centering
\frame{\includegraphics[width=.9\textwidth,height=0.5\paperheight]{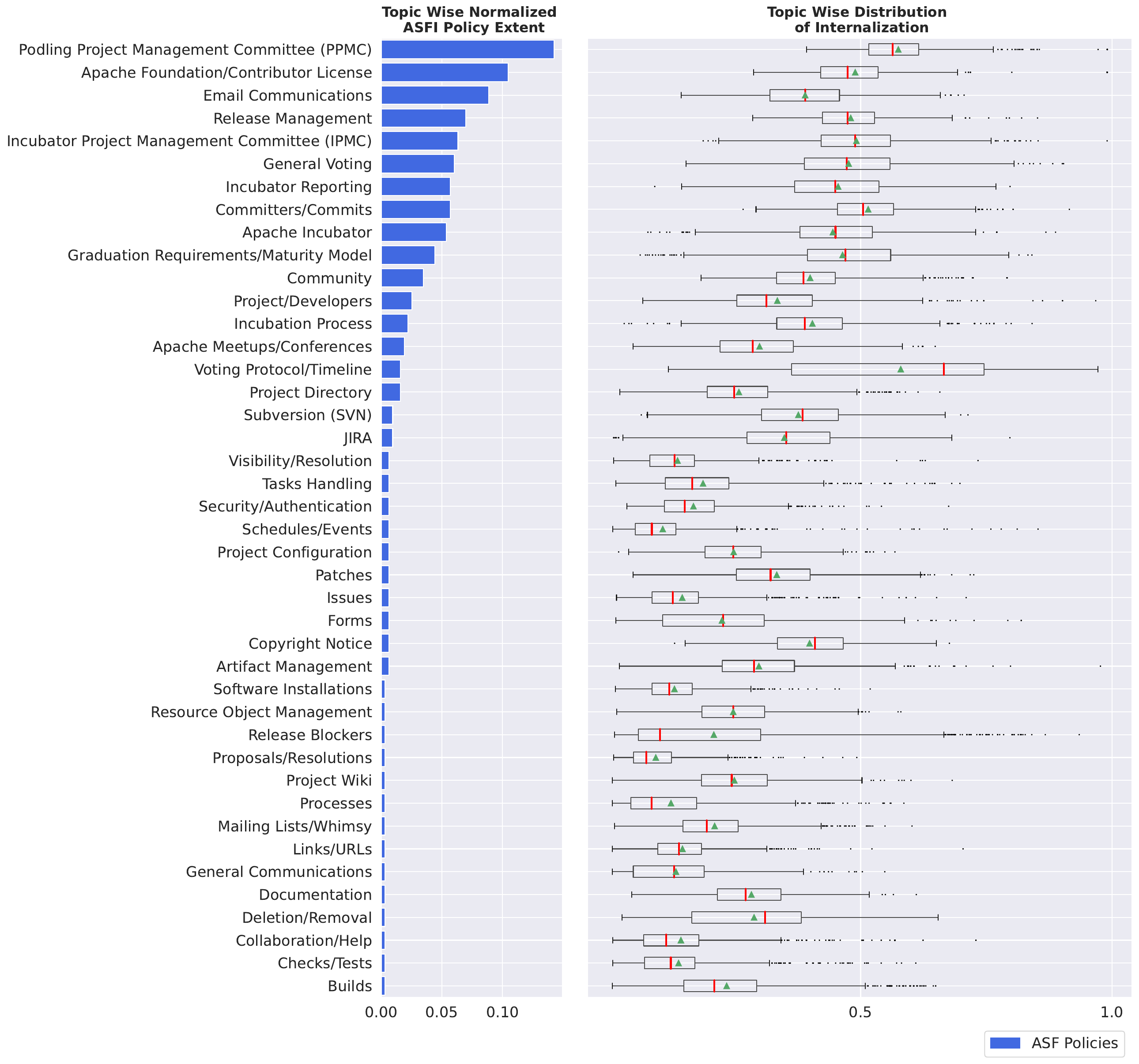}}
\caption{Left: Distribution of ASFI policy extent across governance topics. Right: Distribution of internalization scores within topics. Red and Green markers indicate the median and mean respectively. Internalization is observed to be higher in governance topics which are more regulated.}
\label{fig:RQ2}
\end{figure}

\subsection{RQ2: How do the levels of policy internalization in governed activities relate to ASFI policy extent across different topics?}

To explore RQ2, we additionally examine the distribution of internalization scores of governed activities conditioned on governance topics (Figure ~\ref{fig:RQ2}). Higher mean internalization scores indicate that in a particular topic, the projects' practiced routines are more framed by formalized Incubator policy.  We observe a trend of generally greater internalization with increasing policy extent: a Pearson correlation test between the topic-wise policy extent and mean internalization scores was found to be 0.744 (\emph{p < 0.001}). In other words, areas of governance that receive more attention in formal policy also tend to be enacted by participants in a way close to the policy descriptions. 

\label{section:RQ3}
\subsection{RQ3: How do governed activities and extent of policy internalization relate to the success of projects?}

ASFI strives to build meritocratic communities and assesses projects' performance throughout the incubation time frame. As membership and activity levels undergo constant changes in OSS, we average the monthly measures of active committers, developer emails, and commit activity to capture their sustained levels. The code base variable was represented as the net size of the project repository in terms of overall lines of code (LOC) written by the project while in ASFI. Prior work on ASFI has shown that successful projects tend to graduate early \cite{yin2022open}, so we incorporate the total number of months spent by the project in the Incubator as one of the covariates. To similarly adapt our governance measures, we represent governed activity through the total number of routine activities observed in a project during incubation, across the mailing list. The overall policy internalization along a governance topic every project was similarly evaluated, by averaging the scores across all the governed activities. The resulting number of predictors was 89, including five covariates and the two distinct governance measures across each of the 42 topics. Six projects were dropped as their commit history was unmeasurable through our metrics tool, leading to 208 observations. 

Certain project mailing lists did not reflect governed activity under some of the topics, making the \emph{governed activity} of that topic equal to 0. There are 54 projects with 0 observed \emph{governed activity} in at least one topic. Rather than dropping those observations entirely, we retained them in a way that minimizes information added to the system through the imputation procedure but allows us to retain the information in the non-missing variables: unmeasured internalization scores were filled through iterative round-robin imputing supported by the Python package Sklearn. This method of imputation, a pythonic implementation of MICE \cite{van2011mice} is unbiased relative to other choices we could have made, such as assigning 0.\footnote{The model was run with list-wise deletion for all projects with at least 1 missing value and the only difference was a change of the sign for the number of commits, an effect for which we have low confidence in all our models. We further repeated the analysis without 5 topics with more than 10\% missing entries. We did not observe major changes in effects (size, direction) and significance.}

Project-level covariates (committers, emails, codebase, and commit activity), as well as governed activity for every topic, were log-scaled to address skew as well as to facilitate comparison along the scale of different projects. Subsequently, all variables were standardized through z-score standardization. We then addressed multi-collinearity by removing all variables with Variance Inflation Factors $>$ 5. We then performed a logistic LASSO-based variable selection over 5-fold cross-validation and hyperparameter tuning over the log loss. After multicollinearity tests and variable selection, we have a reduced set of 11 significant predictors. 

We construct nested linear regressions, whereby we fit four models to assess the contribution from different groups of variables (Table. \ref{RQ3}). These are the "baseline" model with only covariates as predictors (M1), a second model adding topical governed activity variables to the baseline (M2), a third model adding only policy internalization variables to the baseline (M3), and the final full model including all three groups of variables: baseline covariates, governance activity, and policy internalization measures (M4). For every model, we additionally checked for outlier influence using Cook's distance and found no data points with extreme leverage ($ D > 1$). The assumptions of log odds linearity were validated using the Box-Tidwell test, whereby no interaction terms $ x*log(x) $ were found significant. We observe that the predictive efficiency and fit of the models improve with step-wise addition of governance variables, a reassuring sign of valid model construction across the three types of variables. The full variable model M4 was found to be the most parsimonious ($\Delta$AIC $=23.05$ with second-best model)  with goodness of fit at 0.648 (Tjur's psuedo-R\textsuperscript{2}). Further, it showed a weighted F1 score and accuracy of 93.6\% and 93.7\% respectively. We hereby report our findings based on M4.

\begin{table}[!htbp]
\caption{\small{Summary RQ3: Binomial(Logit) GLM regression of project governance against Graduation/Retirement} }
\centering
\label{tab:RQ3}
\resizebox{0.99\textwidth}{!}{
\renewcommand{\arraystretch}{2}
\begin{tabular}{lllllllll}
\toprule
& \hfil \textbf{Covariates} & & \hfil \textbf{Covariates and Governed Activity} & & \hfil \textbf{Covariates and Internalization} & & \hfil \textbf{All}  & \\ \midrule
Predictor &  \hfil Coefficient & \hfil\textbf{p} &  \hfil Coefficient  & \hfil \textbf{p}  & \hfil Coefficient  & \hfil \textbf{p}  & \hfil Coefficient & \hfil \textbf{p}  \\ \midrule
Intercept & \hfil 2.490  &  0.000 & \hfil 3.032  &   0.000  & \hfil 3.252 & 0.000 &  \hfil 4.427 & 0.000   \\ 
Committers\footnotemark[2] & \hfil  0.077 & 0.874 & \hfil -0.018  &  0.973   &  \hfil  --0.3074 & 0.637 &  \hfil 0.127 & 0.875  \\ 
Commits\footnotemark[2]  & \hfil 0.705  &  0.140 & \hfil 0.615  &   0.243  &  \hfil  0.772 & 0.195 & \hfil 0.197 & 0.793   \\ 
Developer Emails\footnotemark[2] & \hfil 0.807  &  \textbf{0.016} & \hfil 1.069  &   \textbf{0.020}  &  \hfil  1.000 & \textbf{0.020} & \hfil 1.188 & 0.079   \\ 
Incubation time\footnotemark[1]  & \hfil -0.518  &  \textbf{0.011} & \hfil -0.181  &   0.555  &  \hfil  -0.799 & \textbf{0.004} & \hfil -0.334 & 0.420   \\ 
\midrule
Incubator Reporting\footnotemark[2]   & \hfil   &   & \hfil -1.210  &   \textbf{0.011}  & \hfil   &  &  \hfil -1.827 & \textbf{0.002}   \\ 
Patches\footnotemark[2]   & \hfil   &   & \hfil 0.688  &   \textbf{0.011}  &  \hfil &  & \hfil 1.009 & \textbf{0.009}   \\ 
\midrule
Project Configuration\footnotemark[1]  & \hfil   &   & \hfil  &    &  \hfil  0.765 & \textbf{0.002} & \hfil 0.623 & \textbf{0.043}  \\ 
Task Handling\footnotemark[1]  & \hfil  &   & \hfil &    &  \hfil  -0.511 & 0.054 & \hfil -0.609 & 0.084  \\ 
Project Wiki\footnotemark[1]   & \hfil &  & \hfil  &    &  \hfil  -0.720 & \textbf{0.032} & \hfil -1.417 & \textbf{0.005}  \\ 
Voting Protocol/Timeline\footnotemark[1] & \hfil  &  & \hfil  &   &  \hfil  0.428 & 0.129 & \hfil 0.933 & \textbf{0.013}  \\ 
Graduation Requirements/Maturity Mode\footnotemark[1]  & \hfil  &  & \hfil  &   &  \hfil  0.898 & \textbf{0.001} & \hfil 1.058 & \textbf{0.002}   \\ 
\midrule
Observations: 208 &   \hfil R\textsuperscript{2} (Tjur):  & 0.258 &  \hfil R\textsuperscript{2} (Tjur):  & 0.360 & \hfil R\textsuperscript{2} (Tjur):  & 0.486 &  \hfil R\textsuperscript{2} (Tjur):   & 0.648 \\
\hfil &   \hfil AIC:  &  139.91 & \hfil AIC:  & 124.96 & \hfil AIC: & 113.34  & \hfil  AIC:  & 90.29 \\
\bottomrule
\end{tabular}}
\renewcommand{\arraystretch}{1}
\label{RQ3}
\end{table}
\footnotetext[1]{Standardized variables}
\footnotetext[2]{Log transformed (base 10) and standardized variable}

Factors that correlate positively with a project's chance of graduating include greater internalization of policies related to "Project configuration", "Graduation requirements/Maturity Model", and "Voting protocol/Timeline." Moreover, projects that govern patch-handling activities i.e. more governed activity in "Patches", are associated with higher graduation odds. On the other hand, factors that correlate negatively with successful graduation include high internalization of "Project Wiki" and a higher volume of governed activity on Incubator reporting. 

We observe that neither governed activity around nor internalization of the five most highly regulated topics (those on committees, licensing, email communications, and releases) predicts project success. In fact, project success seems to be correlated mostly with the internalization of policies that receive little attention in formal policy. This further complements our overall finding that projects do not run how they say they run, to suggest that, formal policies may not present the full picture of how communities govern to sustain themselves.

Our primary analysis is correlational and not causal. This is important to emphasize because our findings for the "Graduation Requirements" topics are probably a spurious but encouraging validity check: it is likely that the act of a project graduating and conducting necessary protocols explains the positive effect of internalization of graduation policies. Similarly, "Project Wiki"  is composed of a policy that is only activated once the Incubator has voted to retire a project. The most likely explanation for its negative effect is that project retirement is causing policy enactment, not the other way around.

To check the robustness and probe some unidirectional interpretations, we perform a post-hoc analysis where we repeat all experiments with a modified policy dataset that excludes these confounding end-of-incubation-related policies that happen after a determination of graduation or retirement has been made. We focus this robustness analysis exclusively on policies that are relevant to the active incubation and growth phase of ASFI projects. Therefore, we removed 34 out of the 234 policies that are generally applicable for projects post-graduation/retirement or only at the terminal stage of incubation (graduation vote, transferring trademarks, or ceremonial protocols of graduation/retirement, etc.). For RQ1 and RQ2, we once again retain the previously observed trend, or lack thereof, between policy extent, governed activity, and internalization. For RQ3, we retain significant effects from three out of the six variables that stood out in our original analysis. These include "Patches" (governed activity), "Incubator Reporting" (governed activity), and "Voting Protocols/Timeline" ("Internalization). As expected, we no longer observe the significant effect associated with 'Graduation requirements' which comprised several policies (now removed) closely related to the graduation event, while 'Project Wiki' which treated post-retirement project wrap-up, was not among the topics inferred from the reduced set of policies.  Lastly, the topic 'Project Configuration' does not exert a significant influence on project outcomes.  Details are provided in the Appendix. ~\ref{AppA}.

\section{Findings}


We find substantial differences between the policy-making attention of the ASFI and community governance across topics. Results from RQ1 (Figure ~\ref{fig:RQ1}) show that overall, policy extent has no significant correlation with the frequency of governed activities observed across topics. Yet through RQ2 (Figure ~\ref{fig:RQ2}), we observe that topics with higher policy extent see greater policy internalization. Therefore, while project governance efforts do not mirror the distribution of policy across governance topics, the internalization of policies is highly correlated with how much formal policy governs that topic.


In RQ3 where we test our governance constructs against project outcomes, we find that neither governed activity around nor policy internalization along the most highly regulated subjects predicts project outcomes. Also, most of the topics correlated with project success are relatively lightly regulated. 

Domain knowledge of the ASF Incubator can help us further contextualize the results from RQ3 (Table \ref{RQ3}). Rules from the ‘Project configuration’ topic oversee the steps and requirements for setting up ASF infrastructure. Higher internalization associated with more successful projects likely indicates that the development team is more experienced in navigating and utilizing ASF's resources. 

Democratic communities and consensus building are encoded in ASF's functioning ('The Apache Way') and are a hallmark of the OSS movement generally. ASF requires project-level voting for approving releases, appointing members to the project PMC, admitting committers, etc. Observance of ASF’s standard voting procedures likely indicates shared understanding and streamlined decision-making. Projects that have high internalization with ASF's policies regarding "Voting protocol/Timeline" are successfully hosting and running those votes according to ASF's policies, and mobilizing community participation. 

We find a large negative relationship between the frequency of activities around "Incubator reporting" and the likelihood of graduation. We further investigate and find that projects generally discuss and work on reports only when they are due, except when they 1. miss a deadline and are assigned a new report date, 2. need to keep working to resolve issues in a submitted report, 3. are struggling and asked to report more often. 

Projects often lag in reporting when their development stalls and the community is struggling. In such a situation, the ASFI intervenes actively and necessitates more efforts to motivate the projects to meet standards and resume compliance with Incubator requirements. Therefore the effect is likely associated with struggling projects and how the Incubator interacts with them. If this interpretation holds, the mechanism for our correlative findings is that an outside factor ("struggling project") is driving more reporting and reduced graduation chances.



\section{Discussion}

Our goal was to investigate the relationship between formal policies overseeing OSS communities and their actual self-organizing tendencies. OSS-supporting foundations create policies to encode their concerns and priorities. ASFI introduces formal hierarchies through various offices and committees to organize traditionally free-form OSS communities. They also include requirements to ensure standards of development and conduct among projects. 

Governed activities or routine operations indicate the extent of community governance.  Structured activities along a governance topic indicate how developers coordinate and conduct the bulk of their activities from the underlying beliefs and current needs. Therefore, more governed activities are expected as a community seeks to structure and routinize more of its operations. 

As communities undergo formalization, their governance may be expected to reflect their overarching policy focus. The conventional perception of OSS formalization anticipates more institutional formalities and obligations (Section. \ref{section:RRQ}).  This may be observed as increasing community attention on domains on which ASFI sets more rules, and ensuing routine activity from such structuring.  RQ1 tests whether the attention of community governance aligns with that of formal policies across shared governance domains.


While governed activities reflect the extent of community governance across topics, we are also interested in how communities align formal rules and actual governance behaviors. In their efforts to structure activities, projects may choose formal policies, implement their own norms or a combination of both (Section. \ref{section:orgtheory}).  RQ2 further examines if the extent of formal regulation is related to how community governance integrates them, as observable through the policy internalization of governed activities. 


Our results from RQ1(Figure ~\ref{fig:RQ1}) indicate that the extent of ASF's regulation does not, in general, seem to proportionally increase the intensity of "on-the-ground" governed operations. At the same time, our findings from RQ2 (Figure ~\ref{fig:RQ2}) suggest that through extensive policy-making along specific concerns, ASFI succeeds in using policy to orient community governance, which shows up through policy internalization in governed activity along domains with more extensively defined policies. 

We reconcile the implications of the two approaches to understanding formalization. RQ1 dwells on convergence/divergence in ASFI/community effective governance efforts, i.e. formulating, establishing, and implementing rules and norms to structure activities. Meanwhile, RQ2 examines to what extent community governance incorporates ASFI’s formal policies: literally how much communities internalize formal policy’s framing of a governance issue. The positive correlation between internalization and policy extent likely indicates that certain governance topics that are extensively codified considerably structure governed activity. Yet results from RQ1 indicate that highly formalized governance topics elicit relatively less or no more governance effort from communities as compared to those where fewer formal rules exist. In fact, in several crucial topics with limited regulation, projects exercise substantial governance efforts to sustain The takeaway is that the effect of more formalization in policy seems to be reflected less in the volume of governance activity it spurs, and more in how closely that activity hews with prescribed standards.

The ASFI's policy coverage is largely administrative, and it outlines appropriate protocols for governance concerns it deems important. Consequently, when projects engage in highly regulated domains, they respect and internalize such specifications.  Therefore, while the focus of policy-making may not be reflected in the regular governance concerns of developers, policies still act  as a layer of fundamental governance that is seamlessly integrated into communities. Simply put, developers respect policies that are evidently important and extensively specified, but they are also faced with other concerns beyond those where ASF largely institutes policies. 

The ASFI's policies show relatively less attention to the technical aspects that constitute communities' main governance activities (issues/patches, artifacts, etc.), suggesting that the foundation defers to the discretion and objectives of developers on these subjects. The generally lower policy internalization along core development concerns may be also explained by the fact that technical regulations in ASF are few and rather basic guidelines and expectations than specific conditions. We hence see considerable governed activity along some of these (‘issues’/’patches’/’builds’), reflecting efforts to coordinate fluid communities, channel their contributions, adapt to emerging technology, and meet release targets. 

RQ3 examines the association of self-governance and internalization of foundation policies, with the objective success of projects (Table \ref{RQ3}). It is based on the implicit assumption that projects will perform governance and adopt policies in a manner that helps them attain their objective, which is to graduate from the Incubator. The Incubator assesses projects based on the diversification of the community, the capability to produce compliant software and consistent releases. Interestingly governance behavior around the more highly regulated governance topics does not stand out as significant discriminants between graduated and retired projects.  

Foundation policies may play a role in furthering development, facilitating coordination, and consensus among communities, as analyses showed positive associations between internalization of voting and infrastructure use protocols and odds of graduation. We also find some evidence that community initiative in less regulated governance areas supports project sustainability. Projects that coordinate submission and incorporation of patches more often are both building their community and improving their product, making them more likely to graduate. Such projects were likely able to step up to the limited explicit technical governance to institute their own routines to sustain development.

We have one significant finding around a highly regulated topic: Incubator reporting.  We found a negative association between levels of governed activity around Incubator reporting and the odds of graduation. Reporting to the Apache Incubator is intended to motivate project performance as well as track their progress \cite{yin2021apache}. Therefore, it is interesting that more formalization is associated with a reduced likelihood of graduation for a highly regulated topic. We further explain that this effect from Incubator reporting likely does not imply a straightforward causal relation between formalization and success. It also presents a delicate situation for already struggling projects as they are necessitated to focus their governance more towards the priorities of formal policy. This has sometimes proven to be especially burdensome for small projects. Apache Gossip is such an example, where the small community struggled with the overhead of implementing the regular reporting protocols set by the ASFI and was eventually retired. 

All in all, communities are bound by foundation requirements, especially in domains that elicit a greater volume of formalization. At the same time, their actual governance concentrates on aspects distinct from the ones in which ASFI regulates the most. Importantly, we find limited support for the argument that projects should embrace formalization, be it in terms of aligning governance focus or internalizing policies in more regulated topics, in order to successfully realize their objectives. Therefore, written formal policies from OSS communities may not be a comprehensive account for how their actual governance unfolds.

\subsection{Recommendations}
   
Our findings may carry certain implications for community members in the ASF, or the OSS ecosystem more generally. For example, since policy internalization around project configuration, and voting seem to correlate with project graduation, more formal policy (or informal attention) to these topics may help projects succeed. However, we caution against too literal an interpretation of our findings for practice. Our results may be specific to ASF, and as we have seen, some of these effects are unlikely to have a straightforward causal interpretation.  

Our most responsible recommendation from this research, for practitioners in technology policy in general and OSS in particular, is to be pragmatic about governance, be cognizant of organizational variability and uncertainty, and be watchful but permissive about letting projects drift in their interpretation of policy. This allows volunteer communities to focus on self-regulation, activity, and enforcement of issues that they identify as requiring more clarity or structure. By subsuming policy development processes to community will, foundations are posed to gain a policy design that is informed by low-level daily experiences of contributors, and enjoys the legitimacy of its membership.

\section{Threats and Validity}

The findings presented in this study apply to only the ASF. Future replication across more organizations is hoped to enrich OSS governance research with more general insights. For the purposes of our study, we treat ASFI's standards for graduation as an evaluation of OSS success and viability. The ASFI's stated objectives and standards provide a well-rounded criteria to assess the relation of governance behavior with viable and sustainable communities (Section. \ref{section:RQ3}). It should be noted, however, that projects sometimes have varied reasons for choosing to graduate or discontinuing incubation. Reasons include but are not limited to their sense of cultural fit, or need for ASF's specific portfolio of support servers. Therefore ASFI graduation, while considered a respected and tested model of evaluation, may not generalize to a conclusive metric of OSS success.


Our work is based on large public mailing lists. While these are the central channels for ASFI projects, they also maintain private lists reserved for certain project businesses, including committer voting, etc. These are restricted from public access and are currently beyond our scope. ASFI leadership discourages the use of these lists as much as possible, and they are typically only used for "personnel" matters such as if a contributor is breaking a project's code of conduct or to vote in new committers.

Our study rests on information extracted by semi-supervised learning. The choice of semi-supervised learning was largely motivated by our constructs (Section~\ref{section:method}), the limits of supervised learning, and most importantly to facilitate scalable organizational insight. Unsupervised/semi-supervised methods have known limitations, and are particularly difficult to validate. We tuned the performance of our clustering models utilizing established measures such as clustering validity and NPMI-based topic coherence. However, the very high values of $R^{2}$ that we report for our models are an encouraging sign that these constructs are credibly capturing important aspects of project governance activity.

We named the resulting topic clusters by examining the most frequent words used in them as well as the policies to which they were assigned. This qualitatively distills the essence of the clusters and makes it possible for us to interpret them for purposes of downstream analyses. Therefore, interpretations of topics and associated effects may vary across researchers and leaves room for reification. Through further checks, we find that the topics found in the main and supplementary analysis are largely even if not perfectly identified.

While we used domain-adapted language models wherever available, some tasks like semantic role labeling and semantic scoring were more specialized with limited models and datasets available. Annotating training data consistent with benchmark datasets is complicated for such tasks and limits the scope of the methodology for replicating results. We used models trained on standard benchmark datasets in such cases.

Certain project mailing lists did not reflect governed activity under all of the 42 different governance topics. This could be attributed to the extent of engagement or varied priorities across projects. For example, resource object management routines are likely exclusive to Java-based projects. Moreover, HDBSCAN sets a lower threshold on cluster size  (0.001\% of sample size). This leaves a possibility for the merger of minor routines into clusters representing more general themes, or being classified as outliers.

In  (Section~\ref{section:method}), we explain the computing overheads and limits on input size for transformer-based language models, often truncating broader text context in social interactions \cite{yin2022open}. Moreover, we conduct a granular, performance frame-level analysis of community operations. We encountered a few cases in our dataset where extensive policies with multiple nested or bulleted conditions were truncated, during intermediate preprocessing or parsing stages. Ongoing efforts at supporting longer context windows \cite{tay2020long} for representation learning should expand the scope of language models for discourse analysis. 


  

\section{Conclusion}

Open source software projects join foundations like the Apache Software Foundation despite the "anti-regulatory" tendency of many OSS developers. They do so because the standardized, streamlined governance systems that foundations operate provide clarity, best practices, mentorship, economies of scale, and lower administrative overhead. Yet OSS projects may simultaneously find themselves benefiting from formal structure and/or constrained by it to varying degrees. 

While it is a widely accepted truism that governance in practice often differs from governance in form, demonstrating this at scale, and determining the manner in which formal depictions and ground behavior diverge, has been a challenge. Articulating fundamental questions about governance practices through NLP methods, particularly language modeling, enables us to quantify the governance behavior of projects, including how they govern themselves and internalize formal policy. 

We find that while OSS communities are generally framed by formal regulations, they focus their practical governance efforts in a manner distant from the thrust of formal policy-making. Further, their governance behavior around highly formalized concerns seems to have little bearing on their rates of success and sustainability. What stands out is the adaptability of their governance efforts as well as their internalization of policy around relatively less regulated topics. In conclusion, a comprehensive understanding of peer production, and likely other types of collective action, must account for the institution's formal structure while also measuring how such structuration is received in practice.  

\appendix
\section{Supplementary Analysis} \label{AppA}
\subsection{Policies excluded}
The supplementary analysis looks at community and foundation interaction over governance concerns and policies applicable over active incubation and mentoring. We remove certain categories of policy documents based on subsection headings in the original dataset \cite{Sen_Atkisson_Schweik_2022}. These include "Steps to Retirement", "Deciding to Retire", 'Graduation discussion', "Graduation Approval vote", " The Graduation Process", "Preparing a Charter", "The Recommentation Vote", "Submission of Resolution to the board", "Community Graduation Vote", "Press Releases for new Top Level Projects (TLP)", "Whether to graduate to subproject or to top level project", 'post-graduation tasks', 'Transfering trademarks to the ASF' and "Subproject Acceptance vote". The aforelisted sections span terminal formalities and procedures to initiate and garner community/ASF approval for graduation as well as steps towards formal induction into Apache.  ASFI projects may pursue two modes of post-graduation affiliation, to function as a full fledged independent top level project (TLP) or as a subproject under a TLP. Sections also cover protocols to be observed when a project is being retired. All in all, 34 entires were removed out of the original 234. However, we retain policies which state the goals of ASFI, expected standards, evaluation criteria, and other requirements meant to guide and mentor projects towards success. 

\subsection{Topic Modeling and Correspondence}
We repeated all steps through Sec. \ref{section:method}. The modified policy set produced 328 rules which were used to guide clustering and topic modeling. In order to draw parallels between the topical effects from the two analysis, we tested the extent of topic correspondence between this topic model and the topic model from our primary analyses. We discovered 24 topics among governed activity, of which 22 were a subset of the 42 topics from our primary analysis. A top-N word match between the topics produced by the two models found 82.5\% overlap, while the topical assignment of rules showed a correlation of 0.76 with the primary topic model. 

\subsection{Results}
For RQ1, the correlation between the distributions of policy extent and governed activity was found to be 0.18 ($p=0.41$). For RQ2, the correlation between the distribution of policy extent and the mean internalization by topic was found to be 0.69 ($p < 0.001$). These findings are nearly identical to those reported in the main text. The analyses for RQ3 is as below (see Table \ref{tab:sts}). The differences between these findings are those reported in the main text are discussed in the main text.

\begin{table}[!htbp] 
\caption{\small{Summary: Binomial(Logit) GLM regression of project governance against Graduation/Retirement} }
\centering
\label{tab:sts}
\resizebox{0.95\textwidth}{!}{
\renewcommand{\arraystretch}{2}
\begin{tabular}{lllllllll}
\toprule
& \hfil \textbf{Covariates} & & \hfil \textbf{Covariates and Governed Activity} & & \hfil \textbf{Covariates and Internalization} & & \hfil \textbf{All}  & \\ \midrule
Predictor &  \hfil Coefficient  & \hfil \textbf{p} &  \hfil Coefficient  &  \hfil \textbf{p} & \hfil Coefficient  &  \hfil \textbf{p} & \hfil Coefficient  & \hfil \textbf{p}  \\ \midrule
Intercept & \hfil 2.490  &  0.000 & \hfil 3.186  &   0.000  &  \hfil 2.708 &   0.000  & \hfil 3.353 & 0.000   \\ 
Committers\footnotemark[2] & \hfil  0.077 & 0.874 & \hfil -0.027  &  0.960   &  \hfil 0.127  &   0.819 & \hfil 0.1745 & 0.766  \\ 
Commits\footnotemark[2]  & \hfil 0.705  &  0.140 & \hfil 0.449  &   0.408  &  \hfil 0.209  &  0.705 & \hfil -0.0022 & 0.997   \\ 
Developer Emails\footnotemark[2] & \hfil 0.807  &  \textbf{0.016} & \hfil 0.714  &  0.114  &  \hfil 0.952 &   \textbf{0.021} & \hfil 1.048 & 0.063  \\ 
Incubation time\footnotemark[1]  & \hfil -0.518  &  \textbf{0.011} & \hfil -0.358  &   0.286  &  \hfil -0.468 &   \textbf{0.044}  & \hfil -0.238 & 0.515  \\ \midrule
Incubator Reporting\footnotemark[2]   & \hfil   &   & \hfil -1.707  &   \textbf{0.001}  &  \hfil  &    & \hfil -1.584 & \textbf{0.004}   \\ 
Patches\footnotemark[2]  & \hfil   &   & \hfil 0.797  &   \textbf{0.010}  &  \hfil  &    & \hfil 0.766 & \textbf{0.030}   \\ 
Voting Protocol/Timeline\footnotemark[2]  & \hfil  &  & \hfil 1.012 & \textbf{0.002}  &  \hfil   &   & \hfil 0.6686 & 0.070   \\ \midrule
Voting Protocol/Timeline\footnotemark[1]  & \hfil  &  & \hfil  &   &  \hfil 0.761  &   \textbf{0.004} & \hfil 0.800 & \textbf{0.013}   \\ 
Community\footnotemark[1]  & \hfil  &  & \hfil  &   &  \hfil -0.902  &   \textbf{0.000}  & \hfil -0.5161 & 0.052   \\ \midrule
Observations: 208 &   \hfil R\textsuperscript{2} (Tjur):  & 0.258 &  \hfil R\textsuperscript{2} (Tjur):  & 0.442 & \hfil R\textsuperscript{2} (Tjur):  & 0.413 &  \hfil R\textsuperscript{2} (Tjur):   & 0.508 \\
\hfil &   \hfil AIC:  &  139.91 & \hfil AIC:  & 115.36 & \hfil AIC: & 123.49  & \hfil  AIC:  & 109.76 \\ \bottomrule

\end{tabular}}
\renewcommand{\arraystretch}{1}
\end{table}
\footnotetext[1]{Standardized variables}
\footnotetext[2]{Log transformed (base 10) and standardized variable}

\bibliographystyle{plain}  
\bibliography{main.bib}

\begin{thebibliography}{10}

\bibitem{atkisson2022mentors}
Curtis Atkisson.
\newblock Mentors matter: Association of mentors with project success in the apache software foundation incubator.
\newblock {\em Plos one}, 17(8):e0272764, 2022.

\bibitem{atkisson_mentors_2022}
Curtis Atkisson.
\newblock Mentors matter: {Association} of mentors with project success in the {Apache} {Software} {Foundation} {Incubator}.
\newblock {\em PLOS ONE}, 17(8):e0272764, August 2022.

\bibitem{benkler2006wealth}
Yochai Benkler.
\newblock The wealth of networks: How social production transforms markets and freedom, 2006.

\bibitem{bonial2010propbank}
Claire Bonial, Olga Babko-Malaya, Jinho~D Choi, Jena Hwang, and Martha Palmer.
\newblock Propbank annotation guidelines.
\newblock {\em Center for Computational Language and Education Research, CU-Boulder}, 9, 2010.

\bibitem{Brown_Duguid_1991}
John~Seely Brown and Paul Duguid.
\newblock Organizational learning and communities-of-practice: Toward a unified view of working, learning, and innovation.
\newblock {\em Organization science}, 2(1):40–57, 1991.

\bibitem{Butler_Joyce_Pike_2008}
Brian Butler, Elisabeth Joyce, and Jacqueline Pike.
\newblock Don’t look now, but we’ve created a bureaucracy: the nature and roles of policies and rules in wikipedia.
\newblock In {\em Proceedings of the SIGCHI Conference on Human Factors in Computing Systems}, CHI ’08, page 1101–1110, New York, NY, USA, Apr 2008. Association for Computing Machinery.

\bibitem{cer2017semeval}
Daniel Cer, Mona Diab, Eneko Agirre, Inigo Lopez-Gazpio, and Lucia Specia.
\newblock Semeval-2017 task 1: Semantic textual similarity-multilingual and cross-lingual focused evaluation.
\newblock {\em arXiv preprint arXiv:1708.00055}, 2017.

\bibitem{cer2018universal}
Daniel Cer, Yinfei Yang, Sheng-yi Kong, Nan Hua, Nicole Limtiaco, Rhomni~St John, Noah Constant, Mario Guajardo-Cespedes, Steve Yuan, Chris Tar, et~al.
\newblock Universal sentence encoder.
\newblock {\em arXiv preprint arXiv:1803.11175}, 2018.

\bibitem{Cohen_Bacdayan_1994}
Michael~D. Cohen and Paul Bacdayan.
\newblock Organizational routines are stored as procedural memory: Evidence from a laboratory study.
\newblock {\em Organization science}, 5(4):554–568, 1994.

\bibitem{Cohen_Burkhart_Dosi_Egidi_Marengo_Warglien_Winter_1996}
Michael~D. Cohen, Roger Burkhart, Giovanni Dosi, Massimo Egidi, Luigi Marengo, Massimo Warglien, and Sidney Winter.
\newblock Routines and other recurring action patterns of organizations: contemporary research issues.
\newblock {\em Industrial and corporate change}, 5(3):653–698, 1996.

\bibitem{conneau2019unsupervised}
Alexis Conneau, Kartikay Khandelwal, Naman Goyal, Vishrav Chaudhary, Guillaume Wenzek, Francisco Guzm{\'a}n, Edouard Grave, Myle Ott, Luke Zettlemoyer, and Veselin Stoyanov.
\newblock Unsupervised cross-lingual representation learning at scale.
\newblock {\em arXiv preprint arXiv:1911.02116}, 2019.

\bibitem{Conneau_Kiela_Schwenk_Barrault_Bordes_2017}
Alexis Conneau, Douwe Kiela, Holger Schwenk, Loïc Barrault, and Antoine Bordes.
\newblock Supervised learning of universal sentence representations from natural language inference data.
\newblock In {\em Proceedings of the 2017 Conference on Empirical Methods in Natural Language Processing}, page 670–680, Copenhagen, Denmark, 2017. Association for Computational Linguistics.

\bibitem{crawford1995grammar}
Sue~ES Crawford and Elinor Ostrom.
\newblock A grammar of institutions.
\newblock {\em American political science review}, 89(3):582--600, 1995.

\bibitem{crowston2005coordination}
K~Crowston, K~Wei, Q~Li, Ugur Eseryel, and J~Howison.
\newblock Coordination of free/libre open source software development.
\newblock In {\em ICIS 2005}, 2005.

\bibitem{cyert1963behavioral}
Richard~M Cyert and James~G March.
\newblock A behavioral theory of the firm.
\newblock {\em University of Illinois at Urbana-Champaign's Academy for Entrepreneurial Leadership Historical Research Reference in Entrepreneurship}, 1963.

\bibitem{de2007governance}
Paul~B De~Laat.
\newblock Governance of open source software: state of the art.
\newblock {\em Journal of Management \& Governance}, 11:165--177, 2007.

\bibitem{DeHart-Davis_Chen_Little_2013}
Leisha DeHart-Davis, Jie Chen, and Todd~D. Little.
\newblock Written versus unwritten rules: The role of rule formalization in green tape.
\newblock {\em International Public Management Journal}, 16(3):331–356, Jul 2013.

\bibitem{devlin2018bert}
Jacob Devlin, Ming-Wei Chang, Kenton Lee, and Kristina Toutanova.
\newblock Bert: Pre-training of deep bidirectional transformers for language understanding.
\newblock {\em arXiv preprint arXiv:1810.04805}, 2018.

\bibitem{DiMaggio_Powell_1983}
Paul~J. DiMaggio and Walter~W. Powell.
\newblock The iron cage revisited: Institutional isomorphism and collective rationality in organizational fields.
\newblock {\em American sociological review}, page 147–160, 1983.

\bibitem{Feldman_2003}
M.~S. Feldman.
\newblock A performative perspective on stability and change in organizational routines.
\newblock {\em Industrial and Corporate Change}, 12(4):727–752, Aug 2003.

\bibitem{Feldman_2000}
Martha~S. Feldman.
\newblock Organizational routines as a source of continuous change.
\newblock {\em Organization Science}, 11(6):611–629, Dec 2000.

\bibitem{Feller_Fitzgerald_Hissam_2007}
Joseph Feller, Brian Fitzgerald, and Scott~A. Hissam.
\newblock Nonprofit foundations and their role in community-firm software collaboration.
\newblock {\em MIT Press}, 2007.

\bibitem{apacheself}
The Apache~Software Foundation.
\newblock Guide to successful graduation.

\bibitem{ASF2023}
The Apache~Software Foundation.
\newblock Guide to successful graduation.

\bibitem{Freeman_1972}
Jo~Freeman.
\newblock The tyranny of structurelessness.
\newblock {\em Berkeley Journal of Sociology}, 17:151–164, Jan 1972.

\bibitem{frey2019emergence}
Seth Frey and Robert~W Sumner.
\newblock Emergence of integrated institutions in a large population of self-governing communities.
\newblock {\em PloS one}, 14(7):e0216335, 2019.

\bibitem{frey2022governing}
Seth Frey, Qiankun Zhong, Beril Bulat, William~D Weisman, Caitlyn Liu, Stephen Fujimoto, Hannah~M Wang, and Charles~M Schweik.
\newblock Governing online goods: Maturity and formalization in minecraft, reddit, and world of warcraft communities.
\newblock {\em arXiv preprint arXiv:2202.01317}, 2022.

\bibitem{anthony1984constitution}
Anthony Giddens.
\newblock {\em The constitution of society: Outline of the theory of structuration}.
\newblock Univ of California Press, 1984.

\bibitem{grootendorst2022bertopic}
Maarten Grootendorst.
\newblock Bertopic: Neural topic modeling with a class-based tf-idf procedure.
\newblock {\em arXiv e-prints}, pages arXiv--2203, 2022.

\bibitem{heckman2007emergent}
Robert Heckman, Kevin Crowston, U~Yeliz Eseryel, James Howison, Eileen Allen, and Qing Li.
\newblock Emergent decision-making practices in free/libre open source software (floss) development teams.
\newblock In {\em Open Source Development, Adoption and Innovation: IFIP Working Group 2.13 on Open Source Software, June 11--14, 2007, Limerick, Ireland 3}, pages 71--84. Springer, 2007.

\bibitem{hergueux_follow_2022}
Jérôme Hergueux and Samuel Kessler.
\newblock Follow the {Leader}: {Technical} and {Inspirational} {Leadership} in {Open} {Source} {Software}.
\newblock In {\em {CHI} {Conference} on {Human} {Factors} in {Computing} {Systems}}, pages 1--15, New Orleans LA USA, April 2022. ACM.

\bibitem{Hippel_Krogh_2003}
Eric~von Hippel and Georg~von Krogh.
\newblock Open source software and the “private-collective” innovation model: Issues for organization science.
\newblock {\em Organization science}, 14(2):209–223, 2003.

\bibitem{izquierdo2018role}
Javier Luis~C{\'a}novas Izquierdo and Jordi Cabot.
\newblock The role of foundations in open source projects.
\newblock In {\em Proceedings of the 40th international conference on software engineering: software engineering in society}, pages 3--12, 2018.

\bibitem{Izquierdo_Cabot_2020}
Javier Luis~Cánovas Izquierdo and Jordi Cabot.
\newblock A survey of software foundations in open source.
\newblock {\em arXiv:2005.10063 [cs]}, May 2020.
\newblock arXiv: 2005.10063.

\bibitem{jensen_governance_2010}
Chris Jensen and Walt Scacchi.
\newblock Governance in open source software development projects: {A} comparative multi-level analysis.
\newblock In {\em {IFIP} {International} {Conference} on {Open} {Source} {Systems}}, pages 130--142. Springer, 2010.

\bibitem{jhaver2021decentralizing}
Shagun Jhaver, Seth Frey, and Amy Zhang.
\newblock Decentralizing platform power: A design space of multi-level governance in online social platforms.
\newblock {\em arXiv preprint arXiv:2108.12529}, 2021.

\bibitem{jurafsky2000speech}
Dan Jurafsky.
\newblock {\em Speech \& language processing}.
\newblock Pearson Education India, 2000.

\bibitem{Lammers_2011}
John~C. Lammers.
\newblock How institutions communicate: Institutional messages, institutional logics, and organizational communication.
\newblock {\em Management Communication Quarterly}, 25(1):154–182, 2011.

\bibitem{Lave_Wenger_1991}
Jean Lave and Etienne Wenger.
\newblock {\em Situated learning: Legitimate peripheral participation}.
\newblock Cambridge university press, 1991.

\bibitem{Lee_Cole_2003}
Gwendolyn~K. Lee and Robert~E. Cole.
\newblock From a firm-based to a community-based model of knowledge creation: The case of the linux kernel development.
\newblock {\em Organization science}, 14(6):633–649, 2003.

\bibitem{Levitt_March_1988}
Barbara Levitt and James~G. March.
\newblock Organizational learning.
\newblock {\em Annual review of sociology}, page 319–340, 1988.

\bibitem{li_code_2021}
Renee Li, Pavitthra Pandurangan, Hana Frluckaj, and Laura Dabbish.
\newblock Code of conduct conversations in open source software projects on github.
\newblock {\em Proceedings of the ACM on Human-computer Interaction}, 5(CSCW1):1--31, 2021.
\newblock Publisher: ACM New York, NY, USA.

\bibitem{March_1991}
James~G. March.
\newblock Exploration and exploitation in organizational learning.
\newblock {\em Organization science}, 2(1):71–87, 1991.

\bibitem{march1958organizations}
JG~March and HA~Simon.
\newblock Organizations.
\newblock {\em Wiley}, 1958.

\bibitem{Markus_2007}
M.~Lynne Markus.
\newblock The governance of free/open source software projects: monolithic, multidimensional, or configurational?
\newblock {\em Journal of Management \& Governance}, 11(2):151–163, Jun 2007.

\bibitem{McGinnis_2011}
Michael~D. McGinnis.
\newblock An introduction to iad and the language of the ostrom workshop: A simple guide to a complex framework: Mcginnis: Iad guide.
\newblock {\em Policy Studies Journal}, 39(1):169–183, Feb 2011.

\bibitem{mcginnis1999polycentricity}
Michael~Dean McGinnis.
\newblock {\em Polycentricity and local public economies: Readings from the workshop in political theory and policy analysis}.
\newblock University of Michigan Press, 1999.

\bibitem{mcinnes2017hdbscan}
Leland McInnes, John Healy, and Steve Astels.
\newblock hdbscan: Hierarchical density based clustering.
\newblock {\em J. Open Source Softw.}, 2(11):205, 2017.

\bibitem{meyer1977institutionalized}
John~W Meyer and Brian Rowan.
\newblock Institutionalized organizations: Formal structure as myth and ceremony.
\newblock {\em American journal of sociology}, 83(2):340--363, 1977.

\bibitem{Midha_Bhattacherjee_2012}
Vishal Midha and Anol Bhattacherjee.
\newblock Governance practices and software maintenance: A study of open source projects.
\newblock {\em Decision Support Systems}, 54(1):23–32, 2012.

\bibitem{mikolov2013efficient}
Tomas Mikolov, Kai Chen, Greg Corrado, and Jeffrey Dean.
\newblock Efficient estimation of word representations in vector space.
\newblock {\em arXiv preprint arXiv:1301.3781}, 2013.

\bibitem{Mockus_Fielding_Herbsleb_2002}
Audris Mockus, Roy~T. Fielding, and James~D. Herbsleb.
\newblock Two case studies of open source software development: Apache and mozilla.
\newblock {\em ACM Transactions on Software Engineering and Methodology (TOSEM)}, 11(3):309–346, 2002.

\bibitem{moulavi2014density}
Davoud Moulavi, Pablo~A Jaskowiak, Ricardo~JGB Campello, Arthur Zimek, and Jorg Sander.
\newblock Density-based clustering validation.
\newblock In {\em 14th SIAM International Conference on Data Mining}, pages 839--847. Society for Industrial and Applied Mathematics Publications, 2014.

\bibitem{orr1990talking}
Julian~E Orr.
\newblock {\em Talking about machines: An ethnography of a modern job}.
\newblock Cornell University, 1990.

\bibitem{ostrom2009understanding}
Elinor Ostrom.
\newblock {\em Understanding institutional diversity}.
\newblock Princeton university press, 2009.

\bibitem{ostrom2007going}
Elinor Ostrom, Marco~A Janssen, and John~M Anderies.
\newblock Going beyond panaceas.
\newblock {\em Proceedings of the National Academy of Sciences}, 104(39):15176--15178, 2007.

\bibitem{o2003guarding}
Siobh{\'a}n O’Mahony.
\newblock Guarding the commons: how community managed software projects protect their work.
\newblock {\em Research policy}, 32(7):1179--1198, 2003.

\bibitem{o2007governance}
Siobh{\'a}n O’Mahony.
\newblock The governance of open source initiatives: what does it mean to be community managed?
\newblock {\em Journal of Management \& Governance}, 11:139--150, 2007.

\bibitem{pennington2014glove}
Jeffrey Pennington, Richard Socher, and Christopher~D Manning.
\newblock Glove: Global vectors for word representation.
\newblock In {\em Proceedings of the 2014 conference on empirical methods in natural language processing (EMNLP)}, pages 1532--1543, 2014.

\bibitem{Pentland_2005}
B.~T. Pentland.
\newblock Organizational routines as a unit of analysis.
\newblock {\em Industrial and Corporate Change}, 14(5):793–815, Aug 2005.

\bibitem{Pentland_1995}
Brian~T. Pentland.
\newblock Grammatical models of organizational processes.
\newblock {\em Organization Science}, 6(5):541–556, Oct 1995.

\bibitem{qi2020stanza}
Peng Qi, Yuhao Zhang, Yuhui Zhang, Jason Bolton, and Christopher~D Manning.
\newblock Stanza: A python natural language processing toolkit for many human languages.
\newblock {\em arXiv preprint arXiv:2003.07082}, 2020.

\bibitem{Raymond_1999}
Eric Raymond.
\newblock The cathedral and the bazaar.
\newblock {\em Knowledge, Technology \& Policy}, 12(3):23–49, Sep 1999.

\bibitem{Reimers_Gurevych_2019}
Nils Reimers and Iryna Gurevych.
\newblock Sentence-bert: Sentence embeddings using siamese bert-networks.
\newblock In {\em Proceedings of the 2019 Conference on Empirical Methods in Natural Language Processing and the 9th International Joint Conference on Natural Language Processing (EMNLP-IJCNLP)}, page 3980–3990, Hong Kong, China, 2019. Association for Computational Linguistics.

\bibitem{Röder_Both_Hinneburg_2015}
Michael Röder, Andreas Both, and Alexander Hinneburg.
\newblock Exploring the space of topic coherence measures.
\newblock In {\em Proceedings of the Eighth ACM International Conference on Web Search and Data Mining}, page 399–408, Shanghai China, Feb 2015. ACM.

\bibitem{Sawhney_Prandelli_2000}
Mohanbir Sawhney and Emanuela Prandelli.
\newblock Communities of creation: managing distributed innovation in turbulent markets.
\newblock {\em California management review}, 42(4):24–54, 2000.

\bibitem{schweik2007tragedy}
Charles~M Schweik and Robert English.
\newblock Tragedy of the foss commons? investigating the institutional designs of free/libre and open source software projects.
\newblock {\em First Monday}, 2007.

\bibitem{Schweik_English_2013}
Charles~M. Schweik and Robert English.
\newblock Preliminary steps toward a general theory of internet-based collective-action in digital information commons: Findings from a study of open source software projects.
\newblock {\em International Journal of the Commons}, 7(2):234, Aug 2013.

\bibitem{Schweik_English_2012}
Charles~M. Schweik and Robert~C. English.
\newblock {\em Internet success: a study of open-source software commons}.
\newblock MIT Press, 2012.

\bibitem{scott2005institutional}
W~Richard Scott et~al.
\newblock Institutional theory: Contributing to a theoretical research program.
\newblock {\em Great minds in management: The process of theory development}, 37(2):460--484, 2005.

\bibitem{Sen_Atkisson_Schweik_2022}
Anamika Sen, Curtis Atkisson, and Charlie Schweik.
\newblock Cui bono: Do open source software incubator policies and procedures benefit the projects or the incubator?
\newblock {\em International Journal of the Commons}, 16(1), 2022.

\bibitem{Shah_2006}
Sonali~K. Shah.
\newblock Motivation, governance, and the viability of hybrid forms in open source software development.
\newblock {\em Management science}, 52(7):1000–1014, 2006.

\bibitem{shi2019simple}
Peng Shi and Jimmy Lin.
\newblock Simple bert models for relation extraction and semantic role labeling.
\newblock {\em arXiv preprint arXiv:1904.05255}, 2019.

\bibitem{srivastava_enculturation_2018}
Sameer~B. Srivastava, Amir Goldberg, V.~Govind Manian, and Christopher Potts.
\newblock Enculturation {Trajectories}: {Language}, {Cultural} {Adaptation}, and {Individual} {Outcomes} in {Organizations}.
\newblock {\em Management Science}, 64(3):1348--1364, March 2018.

\bibitem{stewart2006impact}
Katherine~J Stewart and Sanjay Gosain.
\newblock The impact of ideology on effectiveness in open source software development teams.
\newblock {\em Mis Quarterly}, pages 291--314, 2006.

\bibitem{Strang_Meyer_1993}
David Strang and John~W. Meyer.
\newblock Institutional conditions for diffusion.
\newblock {\em Theory and society}, page 487–511, 1993.

\bibitem{10.1145/3540250.3549132}
undefinedtefan St\u{a}nciulescu, Likang Yin, and Vladimir Filkov.
\newblock Code, quality, and process metrics in graduated and retired asfi projects.
\newblock In {\em Proceedings of the 30th ACM Joint European Software Engineering Conference and Symposium on the Foundations of Software Engineering}, ESEC/FSE 2022, page 495–506, New York, NY, USA, 2022. Association for Computing Machinery.

\bibitem{stanciulescu2022code}
Ştefan Stănciulescu, Likang Yin, and Vladimir Filkov.
\newblock Code, quality, and process metrics in graduated and retired asfi projects.
\newblock In {\em Proceedings of the 30th ACM Joint European Software Engineering Conference and Symposium on the Foundations of Software Engineering}, pages 495--506, 2022.

\bibitem{tay2020long}
Yi~Tay, Mostafa Dehghani, Samira Abnar, Yikang Shen, Dara Bahri, Philip Pham, Jinfeng Rao, Liu Yang, Sebastian Ruder, and Donald Metzler.
\newblock Long range arena: A benchmark for efficient transformers.
\newblock {\em arXiv preprint arXiv:2011.04006}, 2020.

\bibitem{thakur2020augmented}
Nandan Thakur, Nils Reimers, Johannes Daxenberger, and Iryna Gurevych.
\newblock Augmented sbert: Data augmentation method for improving bi-encoders for pairwise sentence scoring tasks.
\newblock {\em arXiv preprint arXiv:2010.08240}, 2020.

\bibitem{van2011mice}
Stef Van~Buuren and Karin Groothuis-Oudshoorn.
\newblock mice: Multivariate imputation by chained equations in r.
\newblock {\em Journal of statistical software}, 45:1--67, 2011.

\bibitem{Vasilescu_Filkov_Serebrenik_2015}
Bogdan Vasilescu, Vladimir Filkov, and Alexander Serebrenik.
\newblock Perceptions of diversity on git hub: A user survey.
\newblock In {\em 2015 IEEE/ACM 8th International Workshop on Cooperative and Human Aspects of Software Engineering}, page 50–56. IEEE, 2015.

\bibitem{weick1976educational}
Karl~E Weick.
\newblock Educational organizations as loosely coupled systems.
\newblock {\em Administrative science quarterly}, pages 1--19, 1976.

\bibitem{wieting2015towards}
John Wieting, Mohit Bansal, Kevin Gimpel, and Karen Livescu.
\newblock Towards universal paraphrastic sentence embeddings.
\newblock {\em arXiv e-prints}, pages arXiv--1511, 2015.

\bibitem{wieting2020bilingual}
John Wieting, Graham Neubig, and Taylor Berg-Kirkpatrick.
\newblock A bilingual generative transformer for semantic sentence embedding.
\newblock In {\em Proceedings of the 2020 Conference on Empirical Methods in Natural Language Processing (EMNLP)}, pages 1581--1594, 2020.

\bibitem{winter1982evolutionary}
Sidney~G Winter and Richard~R Nelson.
\newblock An evolutionary theory of economic change.
\newblock {\em University of Illinois at Urbana-Champaign's Academy for Entrepreneurial Leadership Historical Research Reference in Entrepreneurship}, 1982.

\bibitem{wolf2019huggingface}
Thomas Wolf, Lysandre Debut, Victor Sanh, Julien Chaumond, Clement Delangue, Anthony Moi, Pierric Cistac, Tim Rault, R{\'e}mi Louf, Morgan Funtowicz, et~al.
\newblock Huggingface's transformers: State-of-the-art natural language processing.
\newblock {\em arXiv preprint arXiv:1910.03771}, 2019.

\bibitem{yan2023github}
Yibo Yan, Seth Frey, Amy Zhang, Vladimir Filkov, and Likang Yin.
\newblock Github oss governance file dataset.
\newblock {\em arXiv preprint arXiv:2304.00460}, 2023.

\bibitem{yin2022open}
Likang Yin, Mahasweta Chakraborti, Yibo Yan, Charles Schweik, Seth Frey, and Vladimir Filkov.
\newblock Open source software sustainability: Combining institutional analysis and socio-technical networks.
\newblock {\em Proceedings of the ACM on Human-Computer Interaction}, 6(CSCW2):1--23, 2022.

\bibitem{yin2021apache}
Likang Yin, Zhiyuan Zhang, Qi~Xuan, and Vladimir Filkov.
\newblock Apache software foundation incubator project sustainability dataset.
\newblock In {\em 2021 IEEE/ACM 18th International Conference on Mining Software Repositories (MSR)}, pages 595--599. IEEE, 2021.

\bibitem{vrehuuvrek2011gensim}
Radim Řehůřek, Petr Sojka, et~al.
\newblock Gensim—statistical semantics in python.
\newblock {\em Retrieved from genism. org}, 2011.

\end{thebibliography}

\end{document}